\documentclass[onecolumn,pre,floatfix,10pt]{revtex4}%
\usepackage{amssymb}
\usepackage{amsmath}
\usepackage{graphicx}
\usepackage{amstext}
\usepackage{epsfig}
\usepackage{epstopdf}
\usepackage{color,graphics}
\usepackage[font=scriptsize, justification=centerlast]{caption}
\usepackage{amsfonts}%
\providecommand{\U}[1]{\protect \rule{.1in}{.1in}}

\begin{document}
\title{Linear and nonlinear waves in quantum plasmas with arbitrary degeneracy of electrons}
\author{Fernando Haas}
\affiliation{Physics Institute, Federal University of Rio Grande do Sul, CEP 91501-970, Av.
Bento Gon\c{c}alves 9500, Porto Alegre, RS, Brazil}
\author{Shahzad Mahmood}
\affiliation{Theoretical Physics Division (TPD), PINSTECH, P. O. Nilore Islamabad 44000, Pakistan}

\begin{abstract}
The purpose of this review is to revisit recent results in the literature
where quantum plasmas with arbitrary degeneracy degree are considered. This is
different from a frequent approach, where completely degeneracy is assumed in
dense plasmas. The general reasoning in the reviewed works is to take a
numerical coefficient in from of the Bohm potential term in quantum fluids, in
order to fit the linear waves from quantum kinetic theory in the long
wavelength limit. Moreover, the equation of state for the ideal Fermi gas is
assumed, for arbitrary degeneracy degree. The quantum fluid equations allow
the expedite derivation of weakly nonlinear equations from reductive
perturbation theory. In this way, quantum Korteweg - de Vries and quantum
Zakharov - Kuznetsov equations are derived, together with the conditions for
bright and dark soliton propagation. Quantum ion-acoustic waves in
unmagnetized and magnetized plasmas, together with magnetosonic waves, have
been obtained for arbitrary degeneracy degree. The conditions for the
application of the models, and the physical situations where the mixed dense -
dilute systems exist, have been identified.


\end{abstract}
\maketitle

\section{Introduction}

It is customary in statistical mechanics of Fermions, to assume an equilibrium
distribution function either of a completely degenerate Fermi gas or, in the
extreme opposite, of a Maxwellian gas and therefore entirely neglecting the
Fermi-Dirac statistics in this case. In quantum plasmas, the Fermi-Dirac
character comes from electrons, among other possibilities. The completely
degenerate case applies for $T \ll T_{F}$, where $T$ is the thermodynamic
temperature and $T_{F}$ is the Fermi temperature. Correspondingly, the
Maxwellian approximation is adequate for $T \gg T_{F}$, \textit{viz.} for
dilute, high temperature plasmas. Naturally, it is clear that in some
instances there is not a clear separation between the temperature scales, so
that one is obliged to consider the full Fermi-Dirac distribution for
arbitrary degeneracy degree. The resulting mathematics becomes more involved,
as in the case of the equation of state, now involving polylogarithm functions
(Lewin 1981) and the presence of the chemical potential (Pathria and Beale
2011), besides number density. Intermediate regimes with $T \approx T_{F}$
appear in the plasma generated in inertial confinement (Manfredi and Hurst
2015) or laboratory simulation of high energy-density astrophysical plasmas
which better fit the intermediate quantum-classical situation (Cross
\textit{et al.} 2014). The underlying full Fermi-Dirac distribution has
consequence not only for the equation of state in a macroscopic, fluid-like
system, but also on the quantum diffraction effects present in such
hydrodynamics. The quantum diffraction effects manifest in quantum kinetic
theory in the non-local form of the interaction term in the Wigner-Moyal
equation or, in the quantum fluid equations, in the Bohm potential and
higher-order gradient corrections (Haas 2011).

The purpose of this review is to present recent results in the literature,
where the quantum hydrodynamics of arbitrary degeneracy plasmas was applied
for the propagation of linear and nonlinear waves. The general approach behind
these results is to assume the quantum hydrodynamic and quantum kinetic
approaches should give the same linear dispersion relation in the long
wavelength limit. This amounts to the introduction of a fitting parameter in
the quantum dispersion or Bohm potential term. This is a different approach in
comparison with other methods for quantum plasmas with arbitrary degeneracy.
For instance, Maffa (1993) considered ion acoustic and Langmuir waves
linearizing the Vlasov-Poisson system around a full Fermi-Dirac equilibrium,
disregarding quantum diffraction. Including quantum recoil (or quantum
diffraction), the low-frequency longitudinal plasma response for arbitrary
degeneracy was studied (Mushtaq and Melrose 2009), (Melrose and Mushtaq 2010).
Going from the linear to the nonlinear realm, Eliasson and Shukla (2008) take
the first few moments of the Wigner-Moyal equation in terms of a local
Fermi-Dirac distribution with arbitrary ratio $T/T_{F}$. Dubinov \textit{et
al.} (2010) used a Bernoulli pseudo-potential approach for warm Fermi gases,
see also (Dubinov and Kitaev 2014). The screening potential around a
test charge in a quantum plasma was studied using quantum hydrodynamics (QHD)
in the linear limit with statistical (Fermi) pressure and Bohm potential from
finite temperature kinetic theory (Eliasson \& Akbari, 2016). The authors
showed that the kinetic corrections included in the Bohm potential has a
profound effect on the Thomas scattering cross section in a quantum plasma
with arbitrary degeneracy and their theory is consistent with DFT theory in
the limiting case of complete degeneracy of electrons. 

The manuscript is organized as follows. In section 2 the basic quantum
hydrodynamic equations for non-relativistic ion-acoustic waves in unmagnetized
plasmas are reviewed, along with the equation of state, allowing arbitrary
degeneracy. Linear waves are then investigated and compared to quantum kinetic
results in the long wavelength limit. This amounts to the determination of a
fitting parameter in front of the quantum dispersion term in the fluid
equations. The propagation of weakly nonlinear waves is analyzed by means of
the corresponding Korteweg - de Vries (KdV) equation, admitting arbitrary
degeneracy level. Section 3 performs the same review as section 2, but now
including an external uniform magnetic field. In consequence, the linear
dispersion relation shows diverse new possibilities, for parallel,
perpendicular and oblique wave propagation, besides strong magnetic fields.
The agreement with quantum kinetic theory is shown again with the same
equation of state and numerical coefficient in front of the Bohm potential
term in quantum fluid equations, in the long wavelength limit. The weakly
nonlinear theory in the magnetized case is given in terms of a quantum
Zakharov-Kuznetsov equation, admitting soliton solutions among other
possibilities. Section 4 reviews the propagation of magnetosonic waves in
quantum plasmas with arbitrary degeneracy degree. The weakly nonlinear case is
described in terms of the corresponding KdV equation derived from reductive
perturbation methods. Section 5 address numerical estimates for suitable
experiments in the intermediate dilute-degenerate regime, assuming the same
values of Fermi and thermodynamic temperatures. Section 6 has some conclusions.

\section{Ion-acoustic Waves with Arbitrary Degeneracy Electrons in
Unmagnetized Quantum Plasmas}

In this section, the ion-acoustic waves in non-relativistic unmagnetized
plasmas with arbitrary degeneracy of electrons are investigated. In this
section mainly we review the material from (Haas and Mahmood 2015).

The continuity and momentum equations for non-degenerate ions are
\begin{align}
\frac{\partial n_{i}}{\partial t}+\frac{\partial}{\partial x}(n_{i}u_{i})  &
=0\,,\label{e1}\\
\frac{\partial u_{i}}{\partial t}+u_{i}\frac{\partial u_{i}}{\partial x}  &
=-\frac{e}{m_{i}}\frac{\partial \phi}{\partial x}\text{ }. \label{e2}%
\end{align}
The momentum equation for the inertialess degenerate electron fluid is
\begin{equation}
0=e\frac{\partial \phi}{\partial x}-\frac{1}{n_{e}}\frac{\partial p}{\partial
x}+\frac{\alpha \, \hslash^{2}}{6\,m_{e}}\frac{\partial}{\partial x}\left(
\frac{1}{\sqrt{n_{e}}}\frac{\partial^{2}}{\partial x^{2}}\sqrt{n_{e}}\right)
\,. \label{e3}%
\end{equation}
The Poisson equation is given by
\begin{equation}
\frac{\partial^{2}\phi}{\partial x^{2}}=\frac{e}{\varepsilon_{0}}(n_{e}%
-n_{i})\,. \label{e4}%
\end{equation}
In these equations $\phi$ is the electrostatic potential and $n_{i}$, $n_{e}$
are the ion and electron fluid density respectively while $u_{i}$ is the ion
fluid velocity. Also, $m_{e}$ and $m_{i}$ are the electron and ion masses,
$-e$ is the electronic charge, $\varepsilon_{0}$ the vacuum permittivity and
$\hslash$ the reduced Planck's constant. In electron momentum equation
(\ref{e3}) $p=p(n_{e})$ is the electron's fluid scalar pressure to be
determined from a barotropic equation of state. The last term proportional to
$\hslash^{2}$ in the momentum equation for electrons is the quantum force (or
Bohm potential), responsible for quantum diffraction or quantum tunneling
effects due to the wave like nature of the electrons. The extra dispersion
effects appear through the Bohm potential. However a coefficient $\alpha$
appearing in Bohm potential term is selected in order to fit exactly in the
long wavelength limit the linear dispersion relation obtained from kinetic
theory in 3D Fermi-Dirac distributed electrons (Rubin et al. 1993). The
definition of numerical factor $\alpha$ involves the dimensionality and
temperature of the system (Barker and Ferry 1998) e.g., Gardner (1994) has
obtained a factor $\alpha=1$ for a local Maxwell-Boltzmann equilibrium. In
equilibrium, we have $n_{i0}=n_{e0}$($=n_{0}$ say). The quantum effects of
ions are ignored in the model due to their heavy mass and also their
temperature is ignored.

In order to derive the equation of state for degenerate electrons, consider a
local quasi-equilibrium Fermi-Dirac particle distribution function
$f=f(\mathbf{v},\mathbf{r},t)$ (Pathria and Beale 2011), given by
\begin{equation}
f(\mathbf{v},\mathbf{r},t)=\frac{\mathcal{A}}{1+e^{\beta(\varepsilon-\mu)}}\,,
\label{e5}%
\end{equation}
where $\beta=1/(\kappa_{B}T)$, $\varepsilon=m_{e}v^{2}/2,v=|\mathbf{v}|$ and
$\mu$ is the chemical potential regarded as a function of position
$\mathbf{r}$ and time $t$. Besides, $\kappa_{B}$ is the Boltzmann constant,
$T$ is the (constant) thermodynamic electron's temperature and $\mathbf{v}$ is
the velocity in phase space. Here, $A$ is chosen to ensure the normalization
$\int \!fd^{3}v=n_{e}$, which gives
\begin{equation}
\mathcal{A}=-\, \frac{n_{e}}{\mathrm{Li}_{3/2}(-e^{\beta \mu})}\left(
\frac{\beta m_{e}}{2\pi}\right)  ^{3/2}=2\left(  \frac{m_{e}}{2\pi \hslash
}\right)  ^{3}\,. \label{e8}%
\end{equation}
In the above equation, the last equality appears from the Pauli principle (the
factor two is due to the electron's spin). The quantities $\mu$ and
$\mathcal{A}$ both varies slowly with space and time in the fluid description.
In above equation (\ref{e8}) the polylogarithm function $\mathrm{Li}_{\nu}(z)$
of index $\nu$ is used, which is defined (Lewin 1981) by
\begin{equation}
\mathrm{Li}_{\nu}(\eta)=\frac{1}{\Gamma(\nu)}\int_{0}^{\infty}\frac{s^{\nu-1}%
}{e^{s}/\eta-1}\,ds\,, \label{e80}%
\end{equation}
where $\Gamma(\nu)$ is the Gamma function.

The obtained equilibrium condition of plasma density and chemical potential
from relation (\ref{e8}) is given by%

\begin{equation}
-\, \frac{n_{0}}{\mathrm{Li}_{3/2}(-e^{\beta \mu_{(0)}})}\left(  \frac{\beta
m_{e}}{2\pi}\right)  ^{3/2}=2\left(  \frac{m_{e}}{2\pi \hslash}\right)  ^{3}\,.
\label{e801}%
\end{equation}
where $\mu_{(0)}$ is the equilibrium chemical potential which is used to
differentiate from the standard symbol $\mu_{0}$ used for permeability in free
space. Here $n_{0}$ is the equilibrium electron (ion) number density.

The scalar pressure \ in the absence of drift velocity is
\begin{equation}
p=\frac{m_{e}}{3}\int \!fv^{2}d^{3}v\,. \label{e81}%
\end{equation}
Using Eqs. (\ref{e5}) and (\ref{e8}) in (\ref{e81}), we find
\begin{equation}
p=\frac{n_{e}}{\beta}\frac{\mathrm{Li}_{5/2}(-e^{\beta \mu})}{\mathrm{Li}%
_{3/2}(-e^{\beta \mu})}, \label{e9}%
\end{equation}
which is the barotropic equation of state for electrons.

Now in order to find some limiting cases, we first consider the dilute plasma
limiting case with a low fugacity $e^{\beta \mu}\ll1$ and using $\mathrm{Li}%
_{\nu}(-e^{\beta \mu})\approx-e^{\beta \mu}$, then Eq.(\ref{e9}) gives,
\begin{equation}
p=n_{e}k_{B}T\,, \label{e11}%
\end{equation}
which is the classical isothermal equation of state of Maxwellian distributed electrons.

However, in the dense plasma case with a large fugacity $e^{\beta \mu}\gg1$,
and approximation $\mathrm{Li}_{\nu}(-e^{\beta \mu})\approx-\left(  \beta
\mu \right)  ^{\nu}/\Gamma(\nu+1)$, we obtain from Eq.(\ref{e9})
\begin{equation}
p=\frac{2}{5}\,n_{0}\varepsilon_{F}\, \left(  \frac{n_{e}}{n_{0}}\right)
^{5/3}\,, \label{e12}%
\end{equation}
which is the equation of state for a 3D fully degenerate electron Fermi gas,
expressed in terms of the equilibrium number density $n_{0}$. The Fermi energy
of electron is $\varepsilon_{F}=\kappa_{B}T_{F}=\hbar^{2}(3\pi^{2}n_{0}%
)^{2/3}/(2m_{e})$, which is the same as the equilibrium chemical potential
$\mu_{(0)}$ in this fully degenerate plasma case.

Now using Eqs. (\ref{e8}) and (\ref{e801}), a useful relation between
perturbed and equilibrium electron density is obtained,%
\begin{equation}
n_{e}=n_{0}\, \frac{\mathrm{Li}_{3/2}(-e^{\beta \mu})}{\mathrm{Li}%
_{3/2}(-e^{\beta \mu_{(0)}})}\,, \label{e18}%
\end{equation}
Here it is to be noted that although a 3D FD equilibrium for electron Fermi
gas is assumed for IAW propagation but only one spatial variable $x$ is needed
in the model set of dynamic equations. Our present work is quite different
from Eliasson and Shukla (2008) work in which high frequency (Langmuir) waves
were studied with its application to 1-D laser plasma experiments. In our
case, we are investigating low-frequency (ion-acoustic) by assuming a 3D
isotropic equilibrium and chemical potential is a non-constant i.e., space and
time dependent.

Using equation of state for electrons (\ref{e9}), the momentum equation
(\ref{e3}) for the inertialess electron fluid can be written as
\begin{equation}
0=e\frac{\partial \phi}{\partial x}-\frac{1}{\beta n_{e}}\frac{\mathrm{Li}%
_{3/2}(-e^{\beta \mu})}{\mathrm{Li}_{1/2}(-e^{\beta \mu})}\frac{\partial n_{e}%
}{\partial x}+\frac{\alpha \, \hslash^{2}}{6\,m_{e}}\frac{\partial}{\partial
x}\left(  \frac{1}{\sqrt{n_{e}}}\frac{\partial^{2}}{\partial x^{2}}\sqrt
{n_{e}}\right)  \text{ }, \label{e16}%
\end{equation}
where the property $d\mathrm{Li}_{\nu}(\eta)/d\eta=(1/\eta)\mathrm{Li}_{\nu
-1}(\eta)$ has been used.

The alternative form of above Eq.(\ref{e16}) with minimum numbers of
polylogarithm functions having non-constant arguments is given by,%

\begin{equation}
0=e\frac{\partial \phi}{\partial x}-\frac{1}{\beta n_{0}}\frac{\mathrm{Li}%
_{3/2}(-e^{\beta \mu_{(0)}})}{\mathrm{Li}_{1/2}(-e^{\beta \mu})}\frac{\partial
n_{e}}{\partial x}+\frac{\alpha \, \hslash^{2}}{6\,m_{e}}\frac{\partial
}{\partial x}\left(  \frac{1}{\sqrt{n_{e}}}\frac{\partial^{2}}{\partial x^{2}%
}\sqrt{n_{e}}\right)  \,, \label{e19}%
\end{equation}
where the expression (\ref{e18}) of electron density $n_{e}$ has also been used.

\subsection{Coupling Parameter for Arbitrary Degenerate Electron Plasmas}

In this section, we will find the analytical expression of coupling parameter
for the arbitrary degenerate electrons. It is worth to mentioned that in the
absence of collision, the average electrostatic potential per particle
$\langle U\rangle$ is much smaller than the corresponding average kinetic
energy $\langle K\rangle.$ For any degree of degeneracy, the average potential
energy $\langle U\rangle \approx e^{2}/(4\pi \varepsilon_{0}r_{s})$, where the
Wigner-Seitz radius $r_{s}=\left(  3/4\pi n_{0}\right)  ^{1/3}$ is defined.
However, average kinetic energy $\langle K\rangle=[m_{e}/(2n_{e})]\int
fv^{2}d^{3}v$, gives on equilibrium $\langle K\rangle=(3/2)\kappa_{B}T\left[
\mathrm{Li}_{5/2}(-e^{\beta \mu_{(0)}})/\mathrm{Li}_{3/2}(-e^{\beta \mu_{(0)}%
})\right]  $. Therefore an universal coupling parameter obtained is given by
\begin{equation}
g\equiv \frac{\langle U\rangle}{\langle K\rangle}=\frac{1}{6}\left(  \frac
{4}{3\pi^{2}}\right)  ^{1/3}\frac{e^{2}n_{0}^{1/3}\beta}{\varepsilon_{0}}%
\frac{\mathrm{Li}_{3/2}(-e^{\beta \mu_{(0)}})}{\mathrm{Li}_{5/2}(-e^{\beta
\mu_{(0)}})}\,, \label{e191}%
\end{equation}
using the expression of electron equilibrium density from (\ref{e801}) in
terms of equilibrium fugacity $e^{\beta \mu_{(0)}}$ and temperature $T$, the
coupling parameter can be written as,
\begin{equation}
g=-\frac{\sqrt{\beta m_{e}/2}}{3^{4/3}\pi^{7/6}}\left(  \frac{e^{2}%
}{\varepsilon_{0}\hslash}\right)  \frac{\left[  \mathrm{Li}_{3/2}%
^{2}(-e^{\beta \mu_{(0)}})\right]  ^{2/3}}{\mathrm{Li}_{5/2}(-e^{\beta \mu
_{(0)}})}, \label{e20}%
\end{equation}
which covers both non-degenerate and degenerate limits. Now using the
properties of the polylogarithm function, once can have $g\approx \langle
U\rangle/(\kappa_{B}T)$ for dilute plasma case, while $g\approx \langle
U\rangle/\varepsilon_{F}$, with $\mu_{(0)}\approx \varepsilon_{F}$ for dense plasmas.

In order to find the minimal temperature ($T_{m}$) of electrons satisfying the
low collisionality condition i.e., $g<1$ (Akhiezer et al. 1975) for both
dilute and dense plasma case, we have from Eq.(\ref{e20}) as follows,%

\begin{equation}
\kappa_{B}T>\kappa_{B}T_{m}=\frac{m_{e}}{2\times3^{8/3}\pi^{7/3}}\left(
\frac{e^{2}}{\varepsilon_{0}\hslash}\right)  ^{2}\left[  \frac{\mathrm{Li}%
_{3/2}^{4/3}(-e^{\beta \mu_{(0)}})}{\mathrm{Li}_{5/2}(-e^{\beta \mu_{(0)}}%
)}\right]  ^{2}. \label{e21}%
\end{equation}
The numerical plot of minimal temperature $T_{m}$ vs fugacity ($z=e^{\beta
\mu_{(0)}}$) is shown in Fig. 1, where $T>T_{m}$ condition holds for $g<1$. It
is evident from the Fig.1 that initially from $z\approx0$ or at dilute plasma
regime high values of temperature are need to satisfy the ideality condition
with the increase in plasma density. At $z>9.8$ or at dense plasma regime, the
temperature curve start bending and relatively smaller temperatures are needed
to satisfy the low collisionality conditions, which happens due to Pauli
blocking effect. The maximum of minimal temperature curve occurs approximately
at $z=9.8$, $T=8.5\times10^{4}$ $K$, which corresponds to density
$n_{0}=2.9\times10^{29}$ $m^{-3}$ in dense plasma regime.


\begin{figure}[tbh]
\begin{center}
\includegraphics[width=8cm]{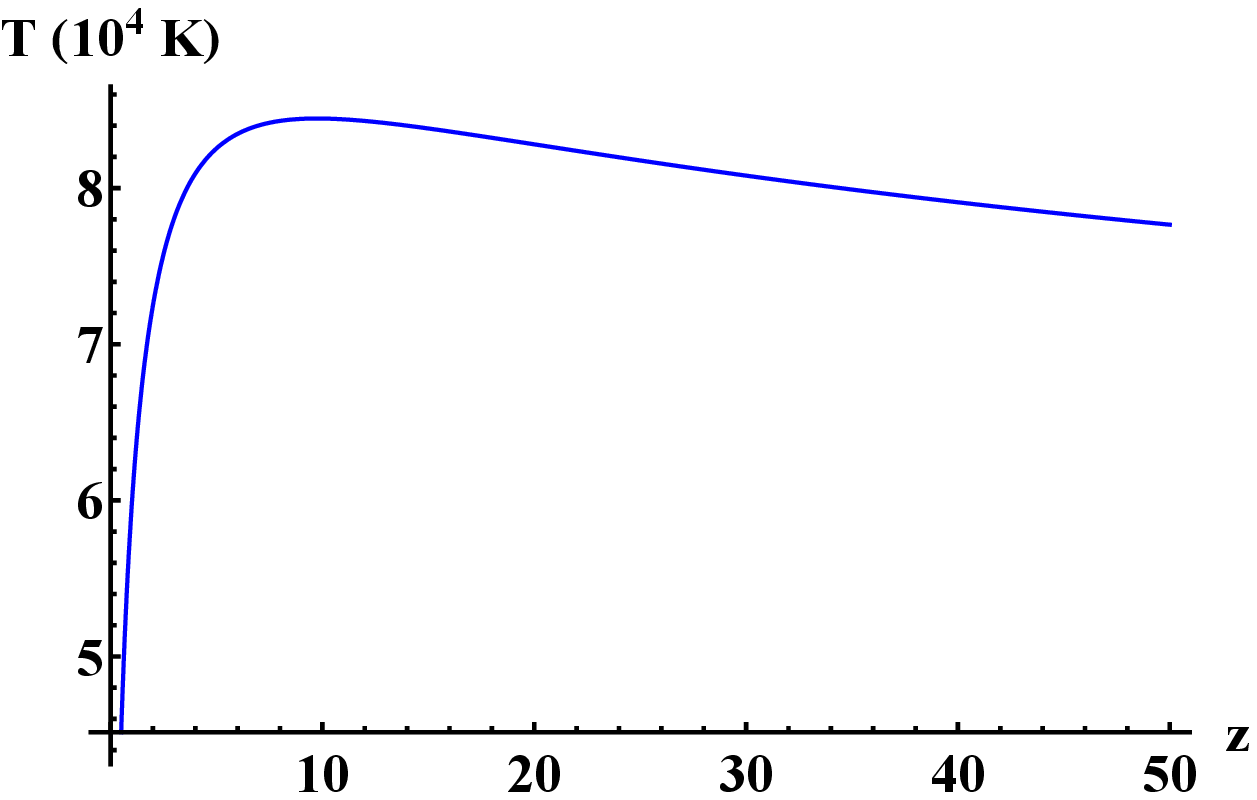}
\end{center}
\caption{Electron temperatures vs equilibrium fugacity $z=e^{\beta \mu_{(0)}}$.
The electron temperature satisfying weak coupling parameter $g<1$ lies above
the curve. }%
\label{figure1}%
\end{figure}

Similarly the universal electron thermal velocity is defined $v_{T}\equiv
\sqrt{2\langle K\rangle/m_{e}}$ and obtained in equilibrium as follows,
\begin{equation}
v_{T}=\left(  \frac{3}{\beta m_{e}}\frac{\mathrm{Li}_{5/2}(-e^{\beta \mu_{(0)}%
})}{\mathrm{Li}_{3/2}(-e^{\beta \mu_{(0)}})}\right)  ^{1/2}\,. \label{e210}%
\end{equation}
Its limiting cases in dilute case, we have $v_{T}\approx \sqrt{3\kappa
_{B}T/m_{e}}$ and in dense plasma case $v_{T}\approx \sqrt{(6/5)\,
\varepsilon_{F}/m_{e}}$. The ion crystallization \ effects (Shukla and
Eliasson 2011; Misra and Shukla 2012) for nondegenerate ions that appear due
to viscoelasticity of the ions fluid in the momentum equation and cause
damping of the ion-acoustic wave are ignored in the model.

\subsection{Linear Wave Analysis}

\subsubsection{Fluid Description}

In order to obtain the dispersion relation of IAW in arbitrary degenerate
electron plasmas, we linearize the set of dynamic equations (\ref{e1}%
)-(\ref{e19}) by assuming the perturbations up to first order and the
dependent variables are described as follows,
\begin{equation}
n_{i}=n_{0}+n_{i1}\,,\quad n_{e}=n_{0}+n_{e1}\,,\quad v_{i}=v_{i1}\,,\quad
\phi=\phi_{1}\,,\quad \mu=\mu_{(0)}+\mu_{1}\,. \label{e211}%
\end{equation}
The expansion of the polylogarithm function up to to first order is given by
\begin{equation}
\mathrm{Li}_{\nu}(-e^{\beta(\mu_{(0)}+\mu_{1})})=\mathrm{Li}_{\nu}%
(-e^{\beta \mu_{(0)}})+\beta \mu_{1}\mathrm{Li}_{\nu-1}(-e^{\beta \mu_{(0)}%
})\text{ }.
\end{equation}
The dispersion relation for IAW is obtained by assuming plane wave
perturbations $\sim \exp[i(kx-\omega t)]$ i.e.,
\begin{equation}
\omega^{2}=\frac{\omega_{pi}^{2}c_{s}^{2}k^{2}\left(  1+\frac{\alpha \,
\hslash^{2}k^{2}}{12\,m_{e}m_{i}c_{s}^{2}}\right)  }{\omega_{pi}^{2}+\left(
1+\frac{\alpha \, \hslash^{2}k^{2}}{12\,m_{e}m_{i}c_{s}^{2}}\right)  c_{s}%
^{2}k^{2}}\,, \label{e22}%
\end{equation}
where a generalized ion-acoustic speed $c_{s}$ is defined as
\begin{equation}
c_{s}=\sqrt{\frac{1}{m_{i}}\left(  \frac{dp}{dn_{e}}\right)  _{0}}=\sqrt
{\frac{1}{\beta m_{i}}\frac{\mathrm{Li}_{3/2}(-e^{\beta \mu_{(0)}}%
)}{\mathrm{Li}_{1/2}(-e^{\beta \mu_{(0)}})}} \label{e23}%
\end{equation}
and $\omega_{pi}=\sqrt{n_{0}e^{2}/(m_{i}\varepsilon_{0})}$ is the ion plasma
frequency. \textbf{Here} $\left(  dp/dn_{e}\right)  _{0}$ is evaluated
at equilibrium. The limiting cases of ion-acoustic speed i.e., in a dilute
plasma (or small fugacity $e^{\beta \mu_{(0)}}\ll1$) $c_{s}\approx c_{sc}%
=\sqrt{\kappa_{B}T/m_{i}}$ is obtained, while in dense plasma case (or
fugacity $e^{\beta \mu_{(0)}}\gg1$) we have $c_{s}=\sqrt{(2/3)\varepsilon
_{F}/m_{i}}$ and it is the same as obtained by (Maafa 1993) for 3D dense plasmas.

In the long wavelength limit $\alpha \, \hslash^{2}k^{4}/(12m_{e}m_{i})\ll
c_{s}^{2}k^{2}\ll \omega_{pi}^{2}$ the dispersion relation (\ref{e22}) IAW
gives $\omega^{2}\approx c_{s}^{2}k^{2}$. However, in short wavelength limit
the dispersion relation just gives an ion oscillations i.e., $\omega
=\omega_{pi}$ and ions are no more shielded by electrons. It happens when the
wavelength becomes comparable or shorter than the electron Debye shielding
length. It can be seen from Eq. (\ref{e23}) that with the increase in fugacity
$e^{\beta \mu_{(0)}}$ (or increase in plasma density) ion-acoustic speed is
also increased. It is interesting to note that taking the square root of both
sides of Eq. (\ref{e22}) is identical to Eq. (4.5) of Ref. (Eliasson and
Shukla, 2010) for completely degenerate plasma case, i.e., $\alpha=1/3$.

\subsubsection{Kinetic Description}

In section, the dispersion relation for IAW will be obtained using kinetic
theory (quantum kinetic) and compare the results already obtained from fluid
model and to set value of numerical coefficient $\alpha$ defined in front of
the quantum force term in the momentum equation of degenerate electrons. The
linear dispersion relation obtained by (Haas 2011) for a Wigner-Poisson system
having cold ion and electron species is given by%

\begin{equation}
1=\frac{\omega_{pi}^{2}}{\omega^{2}}+\frac{\omega_{pe}^{2}}{n_{0}}\int
\frac{f_{0}(\mathbf{v})\,d^{3}v}{(\omega-\mathbf{k}\cdot \mathbf{v}%
)^{2}-\hslash^{2}k^{4}/(4m_{e}^{2})}\,, \label{e24}%
\end{equation}
where $f_{0}(\mathbf{v})$ is the equilibrium electronic Wigner function and
$\omega_{pe}=\sqrt{n_{0}e^{2}/(m_{e}\varepsilon_{0})}$ is the electron plasma
frequency.

Now using the 3D Fermi-Dirac distribution of electrons in equilibrium,
\begin{equation}
f_{0}(\mathbf{v})=\frac{\mathcal{A}}{1+e^{\beta(\varepsilon-\mu_{(0)})}},
\label{e25}%
\end{equation}
the longitudinal response of an electron-ion plasma including the first order
correction from quantum recoil calculated by (Melrose and Mushtaq 2010),
\begin{equation}
1=\frac{\omega_{pi}^{2}}{\omega^{2}}-\frac{\omega_{pi}^{2}}{c_{s}^{2}k^{2}%
}\left \{  1-\frac{m_{e}\left(  \omega^{2}+\hbar^{2}k^{4}/(12\,m_{e}%
^{2})\right)  }{k^{2}\kappa_{B}T}\, \frac{\mathrm{Li}_{-1/2}(-e^{\beta
\mu_{(0)}})}{\mathrm{Li}_{1/2}(-e^{\beta \mu_{(0)}})}\right \}  \label{e26}%
\end{equation}
where $\mathcal{A}$ and $\mu_{(0)}$ are related from Eq. (\ref{e8}) in
equilibrium. Equation (\ref{e26}) follows the Eq.(29) of (Melrose and Mushtaq
2010) but in different notations. The ionic and electronic responses of plasma
appears in the first and second terms on the right-hand side of Eq.(\ref{e26}).

For a low frequency IAW, the static electronic response is taken by putting
$\omega \approx0$ in the last term of Eq. (\ref{e26}), which reduces to%
\begin{equation}
1=\frac{\omega_{pi}^{2}}{\omega^{2}}-\frac{\omega_{pi}^{2}}{c_{s}^{2}k^{2}%
}\left[  1-\frac{\hbar^{2}k^{2}}{12\,m_{e}\kappa_{B}T}\, \frac{\mathrm{Li}%
_{-1/2}(-e^{\beta \mu_{(0)}})}{\mathrm{Li}_{1/2}(-e^{\beta \mu_{(0)}})}\right]
\text{ }. \label{e29}%
\end{equation}
Now solving for wave frequency it gives%
\begin{equation}
\omega^{2}=\frac{\omega_{pi}^{2}c_{s}^{2}k^{2}}{\omega_{pi}^{2}+\left[
1-\frac{\beta^{2}\hbar^{2}\omega_{pe}^{2}}{12}\, \frac{\mathrm{Li}%
_{-1/2}(-e^{\beta \mu_{(0)}})}{\mathrm{Li}_{3/2}(-e^{\beta \mu_{(0)}})}\right]
c_{s}^{2}k^{2}}. \label{e290}%
\end{equation}
The dispersion relation obtained from kinetic theory is valid for wavelengths
larger than the electron shielding length of the system. Therefore, in order
to have comparison with the fluid theory, the dispersion relation of IAW from
(\ref{e290}) is expanded for large wavelengths (or small wave numbers), which gives%

\begin{equation}
\omega^{2}=c_{s}^{2}k^{2}\left[  1+\left \{  -1+\frac{\beta^{2}\hbar^{2}%
\omega_{pe}^{2}}{12}\frac{\mathrm{Li}_{-1/2}(-e^{\beta \mu_{(0)}})}%
{\mathrm{Li}_{3/2}(-e^{\beta \mu_{(0)}})}\right \}  \frac{c_{s}^{2}k^{2}}%
{\omega_{pi}^{2}}\right]  +O(k^{6}). \label{e291}%
\end{equation}
Also expanding the dispersion relation (\ref{e22}) obtained from fluid theory
under large wavelengths, we have%
\begin{equation}
\omega^{2}=c_{s}^{2}k^{2}\left[  1+\left \{  -1+\frac{\alpha \hbar^{2}%
\omega_{pi}^{2}}{12m_{e}m_{i}c_{s}^{2}}\right \}  \frac{c_{s}^{2}k^{2}}%
{\omega_{pi}^{2}}\right]  +O(k^{6}). \label{e292}%
\end{equation}
The obtained dispersion relations from Eqs. (\ref{e291}) and (\ref{e292}) are
equivalent only if, we set%
\begin{equation}
\alpha=\frac{\mathrm{Li}_{3/2}(-e^{\beta \mu_{(0)}})\mathrm{Li}_{-1/2}%
(-e^{\beta \mu_{(0)}})}{\left[  \mathrm{Li}_{1/2}(-e^{\beta \mu_{(0)}})\right]
^{2}}. \label{e293}%
\end{equation}


\begin{figure}[tbh]
\begin{center}
\includegraphics[width=8cm]{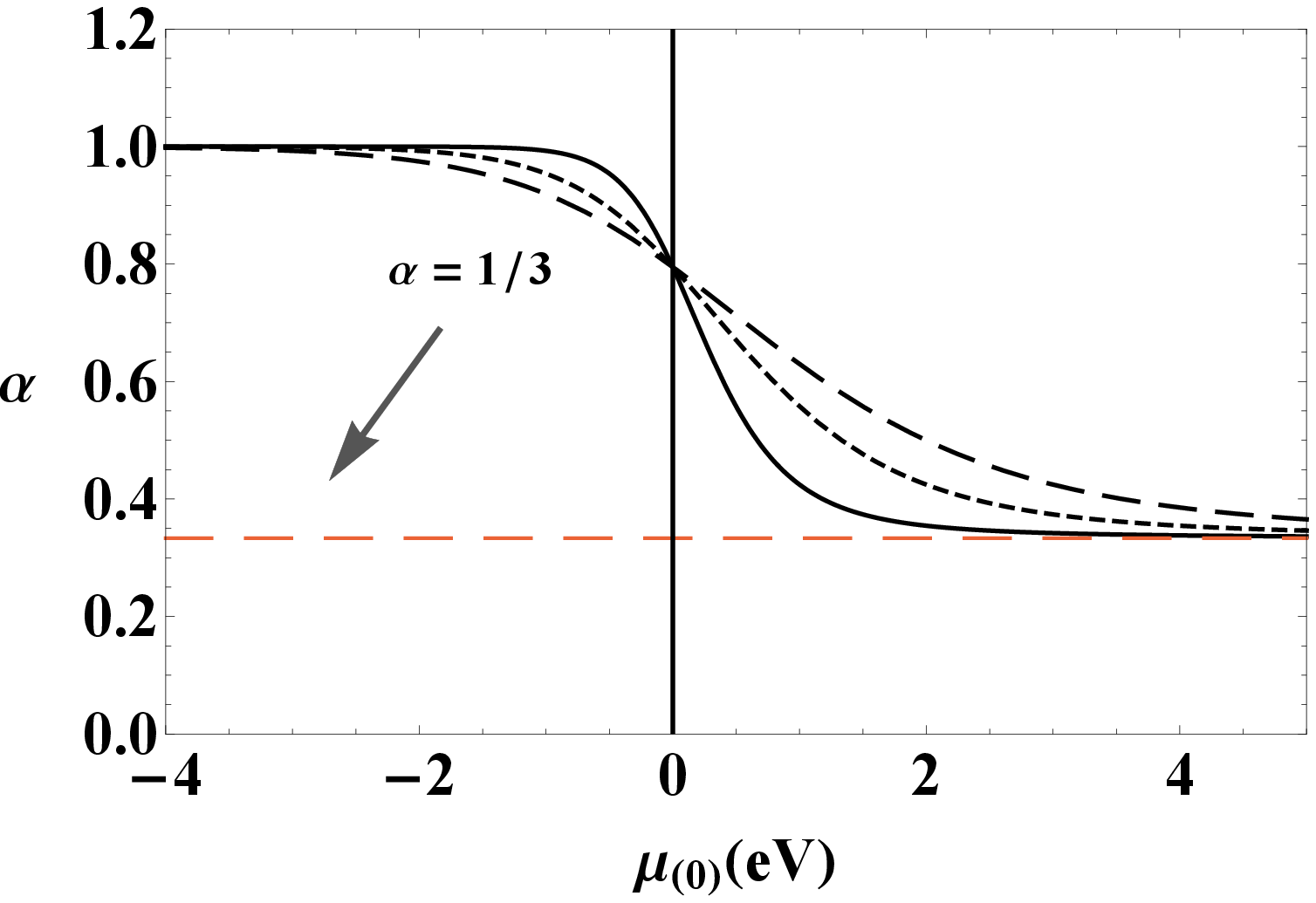}
\end{center}
\caption{Coefficient $\alpha$ from Eq. (\ref{e293}), as a function of the
equilibrium chemical potential $\mu_{(0)}$ is shown at different electron
temperatures i.e., $T=4000\, \mathrm{K}$ (solid line), $T=6000\, \mathrm{K}$
(doted line) and $T=8000\, \mathrm{K}$ (dashed line). }%
\label{figure2}%
\end{figure}

It comes out that the numerical coefficient $\alpha$ in the quantum force term
in the electron momentum equation has to be function of fugacity $\exp
(\beta \mu_{(0)})$ in order to have same dispersion relation from kinetic
theory on a quantum ion-acoustic waves in a 3D Fermi-Dirac equilibrium
electrons. Now if fugacity $z=\exp(\beta \mu_{(0)})$, then chemical potential
in equilibrium $\mu_{(0)}$ can vary from $-\infty$ (classical) to $+\infty$
(dense) plasmas for which $z$ varies from $0$ (dilute) to $\infty$
(ultra-dense) plasmas. Equation (\ref{e293}) gives $\alpha \rightarrow1$ for
$z\rightarrow0$ case, while $\alpha \rightarrow1/3$ for $z\rightarrow \infty$.
Our limiting case result $\alpha \rightarrow1$ agrees for nondegenerate plasma
case with quantum hydrodynamic model for semi-conductor devices derived in
(Gardner 1994), while $\alpha \rightarrow1/3$ agrees with Refs. (Michta et al.
2015; Akbari-Moghanjoughi 2015) in the fully degenerate case. The behavior of
the coefficient $\alpha$ as a function of equilibrium chemical potential
($\mu_{(0)}$) at different temperatures from classical to dense plasma regime
is depicted in Fig.2\textbf{.} It is found that slope of the curve in the
transition region is found to be decreased with the increase in the
temperature of the system. However, the high frequency quantum Langmuir waves
are defined for $\alpha=3$ in order to reproduce the Bohm-Pines dispersion
relation in Ref. (Bohm and Pines 1953).

\subsection{Derivation of Korteweg-de Vries Equation for Ion-acoustic
Solitons}

After performing the linear analysis of quantum ion-acoustic waves in dense
plasma with arbitrary degeneracy of electrons, the above hydrodynamic model
will be considered to find out single pulse nonlinear structures such as solitons.

In order to derive the KdV equation for quantum ion-acoustic waves in
arbitrary degenerate electron plasmas, the set of dynamic equations
(\ref{e1})-(\ref{e4}), can be written in the normalized form as follows,%

\begin{equation}
\frac{\partial \tilde{n}_{i}}{\partial \tilde{t}}+\frac{\partial}{\partial
\tilde{x}}(\tilde{n}_{i}\tilde{u}_{i})=0, \label{e30}%
\end{equation}%
\begin{equation}
\frac{\partial \tilde{u}_{i}}{\partial \tilde{t}}+\tilde{u}_{i}\frac{\partial
}{\partial \tilde{x}}\tilde{u}_{i}=-\frac{\partial \Phi}{\partial \tilde{x}},
\label{e31}%
\end{equation}%
\begin{equation}
0=\frac{\partial \Phi}{\partial \tilde{x}}-\frac{\mathrm{Li}_{1/2}(-e^{\beta
\mu_{(0)}})}{\mathrm{Li}_{1/2}(-e^{\beta \mu})}\frac{\partial \tilde{n}_{e}%
}{\partial \tilde{x}}+\frac{H^{2}}{2}\frac{\partial}{\partial \tilde{x}}\left(
\frac{1}{\sqrt{\tilde{n}_{e}}}\frac{\partial^{2}}{\partial \tilde{x}^{2}}%
\sqrt{\tilde{n}_{e}}\right)  , \label{e32}%
\end{equation}%
\begin{equation}
\frac{\partial^{2}\Phi}{\partial \tilde{x}^{2}}=\tilde{n}_{e}-\tilde{n}_{i},
\label{e33}%
\end{equation}
where a dimensionless quantum diffraction parameter $H$ is
\begin{equation}
H=\frac{\beta \hbar \omega_{_{pe}}}{\sqrt{3}}\left[  \frac{\mathrm{Li}%
_{-1/2}(-e^{\beta \mu_{(0)}})}{\mathrm{Li}_{3/2}(-e^{\beta \mu_{(0)}})}\right]
^{1/2}. \label{e330}%
\end{equation}
The limiting case of quantum diffraction parameter \ for a dilute plasma comes
out to be $H\approx \beta \hbar \omega_{_{pe}}/\sqrt{3}$, while in case of a
fully degenerate plasma $H\approx \hbar \omega_{_{pe}}/(2\varepsilon_{F})$,
respectively. Also, equation (\ref{e18}) in normalized form gives,
\begin{equation}
\tilde{n}_{e}=\frac{\mathrm{Li}_{3/2}(-e^{\beta \mu})}{\mathrm{Li}%
_{3/2}(-e^{\beta \mu_{(0)}})}. \label{e34}%
\end{equation}
The normalization of space, time, ion fluid velocity and electrostatic
potential i.e., $\tilde{x}\rightarrow \omega_{_{pi}}x/c_{s}$, $\tilde
{t}\rightarrow t\omega_{_{pi}}$, $\tilde{u}_{i}\rightarrow u_{i}/c_{s}$ and
$\Phi=e\phi/m_{i}c_{s}^{2}$, respectively, has been defined. Here $c_{s}$ is
the ion-acoustic speed defined in \textbf{equation} (\ref{e23}). The
normalization of electron and ion fluid density is defined as $\tilde{n}%
_{j}=n_{j}/n_{0}$ ($j=e,i$). In further calculations, for simplicity we will
not use the tilde sign on normalized quantities.

In order to find the KdV equation, the stretching of the independent variables
$x$, $t$ is defined as follows (Haas et al. 2003; Mahmood and Haas 2014),%
\[
\xi=\varepsilon^{1/2}(x-V_{0}t)\,,\quad \tau=\varepsilon^{3/2}t,
\]
where $\varepsilon$ is a small expansion parameter and $V_{0}$ is the phase
velocity of the wave to be determined later on. The perturbed quantities can
be expanded in the orders of small expansion parameter as follows,%

\[
n_{j1}=1+\varepsilon n_{j1}+\varepsilon^{2}n_{j2}+...,\text{ (here
}j=e,i\text{)}%
\]%
\[
u_{i}=\varepsilon u_{i1}+\varepsilon^{2}u_{i2}+...,
\]%
\[
\Phi=\varepsilon \Phi_{1}+\varepsilon^{2}\Phi_{2}+...,
\]%
\begin{equation}
\mu=\mu_{(0)}+\varepsilon \mu_{1}+\varepsilon^{2}\mu_{2}+... \label{e37}%
\end{equation}
Now collecting lowest order set of dynamic equations and after solving them,
we get
\begin{equation}
V_{0}=\pm1 \label{e38}%
\end{equation}
which is the normalized phase velocity of the wave. Further more, we set
$V_{0}=1$ without loss of generality. Using above relation (\ref{e38}), the
first order perturbed quantities are related as
\begin{equation}
n_{i1}=u_{i1}=n_{e1}=\Phi_{1}. \label{e39}%
\end{equation}
The next higher order perturbation terms of the set of dynamic equations are
given by
\begin{equation}
\frac{\partial n_{i2}}{\partial \xi}=\frac{\partial n_{i1}}{\partial \tau}%
+\frac{\partial}{\partial \xi}(n_{i1}u_{i1})+\frac{\partial u_{i1}}%
{\partial \tau}+u_{i1}\frac{\partial u_{i1}}{\partial \xi}+\frac{\partial
\Phi_{2}}{\partial \xi}, \label{e40}%
\end{equation}%
\begin{equation}
\frac{\partial n_{e2}}{\partial \xi}=\frac{\partial \Phi_{2}}{\partial \xi
}+\alpha n_{e1}\frac{\partial n_{e1}}{\partial \xi}+\frac{H^{2}}{4}%
\frac{\partial^{3}n_{e1}}{\partial \xi^{3}}. \label{e41}%
\end{equation}
On solving the next higher order Poisson's equation together with Eqs.
(\ref{e38}), (\ref{e40}) and (\ref{e41}), the obtained KdV equation for
quantum ion-acoustic waves in plasmas with arbitrary degeneracy of electrons
\ in terms of $\Phi_{1}$ is given by,%

\begin{equation}
\frac{\partial \Phi_{1}}{\partial \tau}+a\Phi_{1}\frac{\partial \Phi_{1}%
}{\partial \xi}+b\frac{\partial^{3}\Phi_{1}}{\partial \xi^{3}}=0, \label{e42}%
\end{equation}
where the nonlinear and dispersive coefficients $a$ and $b$ i.e.,%
\begin{equation}
a=\frac{3-\alpha}{2}, \label{e43}%
\end{equation}%
\begin{equation}
b=\frac{1}{2}\left(  1-\frac{H^{2}}{4}\right)  , \label{e44}%
\end{equation}
are defined. The limiting case of dilute (classical) plasma for nonlinear and
dispersive coefficients gives $a=1$ and $b=1/2$ respectively in the absence of
Bohm potential, which are same as already derived in Ref. (Davidson 1972) for
the KdV equation in a low density classical plasma. Both nonlinear and
dispersive coefficients have arbitrary degeneracy electrons effects appearing
in $\alpha$ and quantum parameter $H$\textbf{ }respectively.

The traveling wave solution of the KdV equation (\ref{e42}) in terms of single
pulse soliton is given by%

\begin{equation}
\Phi_{1}=D\, \mathrm{sech}^{2}(\eta/W), \label{e45}%
\end{equation}
where $D=3u_{0}/a$ and $W=\sqrt{4b/u_{0}}$ are the amplitude and width of the
soliton, respectively. Also $\eta=\xi-u_{0}\tau$ is the transformed coordinate
in the comoving frame and $u_{0}$ is the speed of the soliton. The potential
hump or dip structure of the quantum IAW soliton depends on the sign of $D$.
The decaying boundary condition have been used for soliton structure. Now
using the relation of stretched coordinates $(\xi,\tau)$ the transformed
coordinate $\eta$ can be described as $\eta=\varepsilon^{1/2}\tilde{\eta}$
where $\tilde{\eta}=x-Vt$ and $V=V_{0}+\varepsilon u_{0}$ is the soliton
velocity in the laboratory frame. The dispersive coefficient described in
Eq.(\ref{e42}) vanishes at $H=2$ and shock formation occurs in the absence of
dispersive effects in the system. In that case, the dispersive effects can be
included through higher order perturbation theory and Kawahara equation is
obtained (Kawahara 1972). The quantum ion-acoustic soliton solution exist only
if $H\neq2$ and dispersive coefficient does not disappear. The nonlinear
coefficient $a$ will always come out to be positive because the numerical
value of parameter $\alpha$ varies from $1$ to $1/3$ from dilute to dense
plasma case. A proper balance between the nonlinearity and dispersion in the
system gives a soliton structure.


\begin{figure}[tbh]
\begin{center}
\includegraphics[width=8cm]{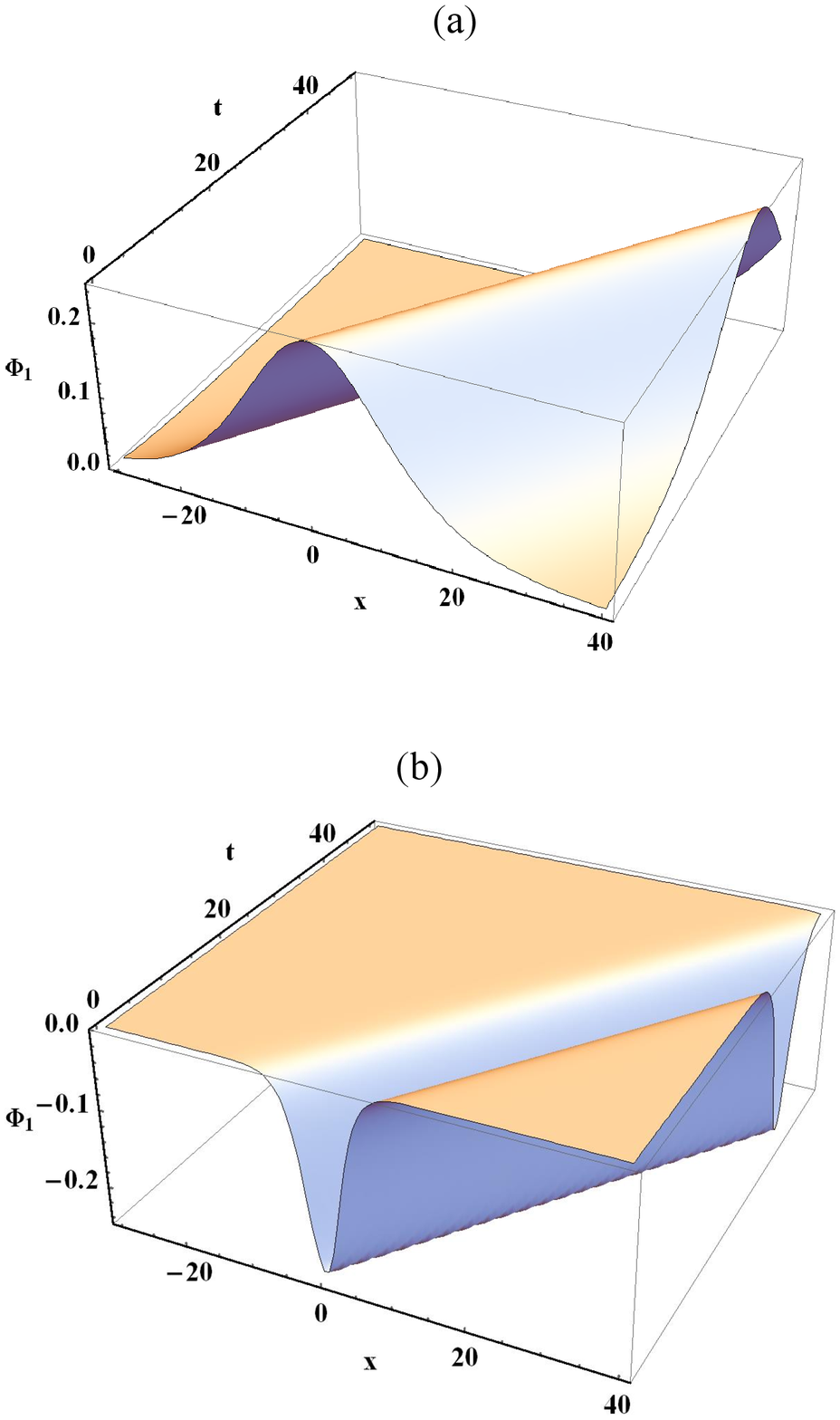}
\end{center}
\caption{(a): Electrostatic potential hump ion-acoustic soliton structure
(moving with supersonic speed in lab frame) is shown from Eq. (\ref{e45}) for
$H<2$ \ case with $\alpha=0.5,$ $H=0.5,$ $\varepsilon=0.1$ and $u_{0}=0.1$.
(b): Electrostatic potential dip ion-acoustic soliton structure (moving with
subsonic speed in lab frame) is shown from Eq. (\ref{e45}) for $H>2$ case with
$\alpha=0.5,$ $H=2.05,$ $\varepsilon=0.1$ and $u_{0}=-0.1$. }%
\label{figure3}%
\end{figure}

In case if $H<2$, then $u_{0}$ must have positive values to give the real
value of soliton width $W$ for which $D>0$ holds and potential hump (bright)
soliton structure is obtained which moves with super-sonic speed in the
forward direction as $V>V_{0}$ . However, when $H>2$, then $u_{0}$ must have
negative values to give the real value of soliton width $W$ for which $D<0$
holds and potential dip (dark) soliton structure is obtained which moves with
sub-sonic speed in the backward direction as $V<V_{0}$ (Belashov and
Vladimirov 2005). Therefore, the formation of both potential hump (bright) and
dip (dark) quantum ion-acoustic soliton are possible in our present model. The
numerical presentation of the bright ($H<2$) and dark ($H>2$) quantum
ion-acoustic solitons obtained from Eq. (\ref{e45}) are shown in Figs. 3a and
3b, respectively. The quantum ion-acoustic potential hump (bright) soliton
structure is obtained for $z=5$, $T=10^{5}$\textrm{K} , which corresponds to
plasma density $n_{0}=3.3\times10^{29}$ $\mathrm{m}^{-3}$ , while quantum
diffraction parameter and electron plasma frequency comes out to be $H=0.64$,
{$\omega_{pe}=3.3\times10^{16}\,$}$\mathrm{rad/s}${, respectively. However,
quantum }ion-acoustic potential dip (dark) soliton structure is obtained for
$z=5$, $T=10^{3}$\textrm{K} , which corresponds to plasma density
$n_{0}=3.5\times10^{26}$ $\mathrm{m}^{-3}$ , while quantum diffraction
parameter and electron plasma frequency comes out to be $H=2.03$ and
{$\omega_{pe}=1.1\times10^{15}\,$}$\mathrm{rad/s}$, respectively. It can be
seen from Eqs.(\ref{e43}) and (\ref{e44}) that for large values of nonlinear
and dispersive coefficients $a$ and $b$ (or in a more dense plasma) the small
amplitude and large width soliton structures are formed. This happens because
under assumption of strong degeneracy, it becomes difficult to accommodate
more Fermions in a localized structure or wave packet.

\subsubsection{Conditions for the existence of bright and dark quantum
ion-acoustic solitons}

In this step, we will look for the physical condition for the propagation of
bright and dark IAW soliton i.e., for $H<2$ or $H>2$ respectively, in a
quantum plasma. Now using the relation (\ref{e801}) of plasma equilibrium
density the dimensionless quantum parameter from Eq. (\ref{e330}) can be
written as%

\begin{equation}
H^{2}=-\frac{1}{3\pi}\left(  \frac{m_{e}}{2\, \pi \kappa_{B}T}\right)
^{1/2}\frac{e^{2}}{\hbar \varepsilon_{0}}\mathrm{Li}_{-1/2}(-e^{\beta \mu_{(0)}%
}). \label{e46}%
\end{equation}
From above equation, the condition for the system temperature at which$H>2$
occurs can be found out as follows%
\begin{equation}
\kappa_{B}T<\frac{m_{e}}{288\, \pi^{3}}\left(  \frac{e^{2}}{\hbar
\varepsilon_{0}}\right)  ^{2}\left[  \mathrm{Li}_{-1/2}(-e^{\beta \mu_{(0)}%
})\right]  ^{2}\text{.} \label{e47}%
\end{equation}
Similarly under the weak coupling condition for an ideal Fermi gas, the upper
bound on quantum diffraction parameter $H$ can be found by using the minimal
temperature in Eq. (\ref{e21}) with Eq. (\ref{e46}), which gives
\begin{equation}
H^{2}<H_{M}^{2}=\left(  \frac{3}{\pi}\right)  ^{1/3}\frac{\mathrm{Li}%
_{-1/2}(-e^{\beta \mu_{(0)}})\mathrm{Li}_{5/2}(-e^{\beta \mu_{(0)}})}{\left[
\mathrm{Li}_{3/2}^{2}(-e^{\beta \mu_{(0)}})\right]  ^{2/3}}. \label{e48}%
\end{equation}
It can be seen numerically that large $H>2$ values fall within the the
strongly coupled regime where the coupling parameter $g$ defined in Eq.
(\ref{e20}) may become near or greater than one.

\section{Ion-acoustic Waves for Magnetized Quantum Plasmas with Arbitrary
Degeneracy of Electrons}

In this section, we will investigate the quantum IAW in a magnetized plasma
with arbitrary degeneracy of electrons. The external magnetic field is assumed
to be directed along x-axis i.e., $\mathbf{B}_{0}=B_{0}\hat{x}$ and the
obliquely propagating electrostatic wave lies in the XY plane i.e.,
$\mathbf{\nabla}=(\partial_{x},\partial_{y},0)$. The ions are considered to be
inertial, while electron quantum fluid is assumed to be inertialess. The
material in this section is a review based on (Haas and Mahmood 2016). The set
of dynamic equations with arbitrary degeneracy of electrons in a magnetized
dense plasma are given below.

The ion continuity and momentum equations are described as follow:
\begin{equation}
\frac{\partial n_{i}}{\partial t}+\frac{\partial}{\partial x}(n_{i}%
u_{ix})+\frac{\partial}{\partial y}(n_{i}u_{iy})=0\,, \label{e51}%
\end{equation}%
\begin{align}
\frac{\partial u_{ix}}{\partial t}+\left(  u_{ix}\frac{\partial}{\partial
x}+u_{iy}\frac{\partial}{\partial y}\right)  u_{ix}  &  =-\frac{e}{m_{i}}%
\frac{\partial \phi}{\partial x}\,,\label{e52}\\
\frac{\partial u_{iy}}{\partial t}+\left(  u_{ix}\frac{\partial}{\partial
x}+u_{iy}\frac{\partial}{\partial y}\right)  u_{iy}  &  =-\frac{e}{m_{i}}%
\frac{\partial \phi}{\partial y}+\omega_{ci}u_{iz}\,,\label{e53}\\
\frac{\partial u_{iz}}{\partial t}+\left(  u_{ix}\frac{\partial}{\partial
x}+u_{iy}\frac{\partial}{\partial y}\right)  u_{iz}  &  =-\omega_{ci}u_{iy}\,.
\label{e54}%
\end{align}
The momentum equation for the inertialess degenerate electrons is
\begin{equation}
0=-\frac{\nabla p}{n_{e}}+e\nabla \phi+\frac{\alpha \hbar^{2}}{6m_{e}}%
\nabla \left[  \frac{1}{\sqrt{n_{e}}}\left(  \frac{\partial^{2}}{\partial
x^{2}}+\frac{\partial^{2}}{\partial y^{2}}\right)  \sqrt{n_{e}}\right]  \,.
\label{e55}%
\end{equation}
The Poisson equation is given by
\begin{equation}
\left(  \frac{\partial^{2}}{\partial x^{2}}+\frac{\partial^{2}}{\partial
y^{2}}\right)  \phi=\frac{e}{\varepsilon_{0}}(n_{e}-n_{i}), \label{e56}%
\end{equation}
here $\phi$ is the electrostatic potential. The electron and ion fluid
densities are represented by $n_{e}$ and $n_{i}$ respectively while
$\mathbf{u}_{i}=(u_{ix},u_{iy},u_{iz})$ is the ion fluid velocity and its
components along $x$, $y$ and $z$ directions. The electron and ion masses are
represented by $m_{e}$ and $m_{i}$ respectively, while $-e$ is the electronic
charge, $\varepsilon_{0}$ is the vacuum permittivity, $\hbar$ is the reduced
Planck constant. The ion cyclotron frequency is$\  \omega_{ci}=eB_{0}/m_{i}$ .
The electron's fluid pressure $p=p(n_{e})$ is specified by a barotropic
equation of state (\ref{e9}) obtained from the moments of a local Fermi-Dirac
distribution function of an ideal Fermi gas. The last term on the right hand
side of the momentum equation (\ref{e55}) for electrons is the quantum force,
which arises from the Bohm potential, giving rise to quantum diffraction or
tunneling effects due to the wave like nature of the charged particles. The
equilibrium is defined as $n_{e0}=n_{i0}\equiv n_{0}$. The numerical factor
$\alpha$\ is defined in the quantum force term of Eq. (\ref{e55}) in order to
fit the kinetic linear dispersion relation in the long wavelength limit, in a
3D Fermi-Dirac equilibrium. The mathematical expression of numerical
coefficient $\alpha$ in terms of polylogarithm function $\mathrm{Li}$ with
fugacity dependence has been derived in Eq. (\ref{e293}) from
finite-temperature quantum kinetic theory for low frequency IAW excitations.
Furthermore, the electron density $n_{e}$\ and chemical potential $\mu$\ are
related through relation (\ref{e18}) in a quasi-Fermi Dirac equilibrium, while
electron equilibrium chemical potential $\mu_{0}$ is related to the
equilibrium density $n_{0}$ through relation (\ref{e801}), respectively.

\subsection{Linear Wave Analysis}

\subsubsection{Fluid Description}

The dispersion relation of electrostatic waves in a magnetized dense plasma
with arbitrary degeneracy of electrons will be discussed in this section. In
order to find the dispersion relation, the set of dynamic ion and electron
fluid equations (\ref{e51})-(\ref{e56}) are linearized up to first order by
defining the dependent variables as follows:%

\begin{equation}
n_{i}=n_{0}+n_{i1}\,,n_{e}=n_{0}+n_{e1}\,,u_{ix}=u_{ix1}\,,u_{iy}%
=u_{iy1}\,,u_{iz}=u_{iz1}\,,\phi=\phi_{1}\, \text{and }\mu=\mu_{(0)}+\mu_{1}.
\end{equation}
By assuming the plane wave perturbations $\sim \exp[i(k_{x}x+k_{y}y-\omega
t)]$, the obtained dispersion relation of electrostatic waves in a magnetized
quantum plasma is
\begin{equation}
1+\chi_{i}(\omega,\mathbf{k})+\chi_{e}(0,\mathbf{k})=0\,, \label{e57}%
\end{equation}
where the ionic and electronic susceptibilities are described by
\begin{align}
\chi_{i}(\omega,\mathbf{k})  &  =-\, \frac{\omega_{pi}^{2}(\omega^{2}%
-\omega_{ci}^{2}\cos^{2}\theta)}{\omega^{2}(\omega^{2}-\omega_{ci}^{2}%
)}\,,\label{e58}\\
\chi_{e}(0,\mathbf{k})  &  =\omega_{pe}^{2}\left[  \frac{1}{m_{e}}\left(
\frac{dp}{dn_{e}}\right)  _{0}k^{2}+\frac{\alpha \hbar^{2}k^{4}}{12m_{e}^{2}%
}\right]  ^{-1}\! \! \!, \label{e59}%
\end{align}
where $\omega_{pj}^{2}=n_{0}e^{2}/(m_{j}\varepsilon_{0})$ for $j=i,e$ and
$\mathbf{k}=k\,(\cos \theta,\sin \theta,0)$. Here $\chi_{e}(0,\mathbf{k})$ is
the static electronic susceptibility due to inertialess electron quantum
fluid. There is no loss of generality in assuming waves propagation in the XY
plane, due to symmetry around the direction of the external magnetic field
i.e., $x$-axis.

The dispersion relation (\ref{e57}) along with Eqs.(\ref{e58}) and (\ref{e59})
gives a quadratic equation for $\omega^{2}$ and its obtained solution is given
by
\begin{equation}
\omega^{2}=\frac{1}{2}\left[  \omega_{0}^{2}+\omega_{ci}^{2}\pm \left \{
(\omega_{0}^{2}+\omega_{ci}^{2})^{2}-4\omega_{0}^{2}\omega_{ci}^{2}\cos
^{2}\theta \right \}  ^{1/2}\right]  \,, \label{e60}%
\end{equation}
and%

\begin{equation}
\omega_{0}^{2}=\frac{c_{s}^{2}k^{2}\left[  1+H^{2}(k\lambda_{D})^{2}/4\right]
}{1+(k\lambda_{D})^{2}+H^{2}(k\lambda_{D})^{4}/4}\,, \label{e61}%
\end{equation}
where $\omega_{0}$ is the wave frequency of quantum IAW in unmagnetized plasma
derived in Eq.(\ref{e22}), $c_{s}$ is the ion-acoustic speed and $H$ is the
quantum diffraction parameters defined in Eqs. (\ref{e23}) and (\ref{e330}%
)\ respectively. Also the generalized screening length of degenerate electrons
is defined as%

\begin{equation}
\lambda_{D}^{2}=\frac{c_{s}^{2}}{\omega_{pi}^{2}}=\frac{\kappa_{B}T}%
{m_{e}\omega_{pe}^{2}}\, \frac{\mathrm{Li}_{3/2}(-e^{\beta \mu_{(0)}}%
)}{\mathrm{Li}_{1/2}(-e^{\beta \mu_{(0)}})}\,. \label{e62}%
\end{equation}
In a low fugacity $e^{\beta \mu_{(0)}}\ll1$ or dilute plasma case, the
ion-acoustic speed and electron Debye length becomes $c_{s}\approx \sqrt
{\kappa_{B}T/m_{i}},\, \lambda_{D}=\sqrt{\kappa_{B}T/(m_{e}\omega_{pe}^{2})}$,
respectively, which are same as in classical plasma. The quantum diffraction
parameter in dilute plasma case is $H\approx \beta \hbar \omega_{pe}/\sqrt{3}$.
However in fully degenerate or dense plasma case i.e., for fugacity
$e^{\beta \mu_{(0)}}\gg1$, the quantum ion-acoustic speed and Thomas-Fermi
screening length comes out to be $c_{s}\approx \sqrt{(2/3)\varepsilon_{F}%
/m_{i}},\, \lambda_{D}=\sqrt{2\varepsilon_{F}/(3m_{e}\omega_{pe}^{2})}$ and
quantum diffraction parameter is $H\approx(1/2)\hbar \omega_{pe}/\varepsilon
_{F}$ where $\mu_{(0)}\approx \varepsilon_{F}=\hbar^{2}(3\pi^{2}n_{0}%
)^{2/3}/(2m_{e})$ is the Fermi energy. The obtained dispersion relation
(\ref{e60}) of electrostatic waves in a magnetized quantum plasma with
arbitrary degeneracy of electrons gives the same dispersion relation for
classical magnetized plasma (Stringer 1963; Witt and Lotko 1983) in the
limiting case of dilute plasmas.

The two symmetrical roots with ``positive" and ``negative" signs of Eq.
(\ref{e60}) corresponds to the fast (ion cyclotron branch) and slow
(ion-acoustic branch) electrostatic waves, respectively, in a
magnetized quantum plasma because electrons stream along the magnetic field
lines to neutralize the ions \  [Bernhardt et al., 2009]. For
almost perpendicular wave propagation to the magnetic field, there exist also
the lower hybrid wave (LHW) which involves the electron inertia via
polarization drift and it is out of scope to our present research work. The
dependence of arbitrary degeneracy of electrons {for arbitrary wave
propagation angle $\theta$ }is not clearly evident from Eq. (\ref{e60}) due to
its complexity. However, in the presence of strong quantum effects
{$\omega_{0}^{2}\gg \omega_{ci}^{2}$, the fast mode gives $\omega^{2}%
\approx \omega_{0}^{2}+\omega_{ci}^{2}\sin^{2}\theta$ and slow mode becomes
$\omega^{2}\approx \omega_{ci}^{2}\cos^{2}\theta$. It means the fast
electrostatic wave frequency has angular dependence as well as quantum
correction, while the slow electrostatic wave\ frequency is only angle
dependent and it is not influenced by quantum effects. In order to have
validity of our model, we consider the example solid density hydrogen
laboratory plasma parameters $T=10^{6}\, \mathrm{K},\,n_{0}=5\times10^{30}\,
\mathrm{m}^{-3}$ in ambient magnetic field $B_{0}=10^{3}\, \mathrm{T}$ and
wave-number $k=2\pi \times10^{9}\, \mathrm{m}^{-1}$, which gives the ion
cyclotron frequency $\omega_{ci}=9.58\times10^{10}\, \mathrm{rad/s}$} and{
$\omega_{0}=6.50\times10^{14}\, \mathrm{rad/s}\gg \omega_{ci}$.}

Some significant limiting cases of Eq. (\ref{e60}) are described below:

\begin{enumerate}
\item[(i)] \textbf{Wave propagation in the parallel direction to the magnetic
field}: In case of wave propagation in the parallel direction to the magnetic
field i.e., $\theta=0$ then two symmetrical roots of the Eq. (\ref{e60}) are
$\omega=\omega_{ci}$ (ion cyclotron frequency) and $\omega=\omega_{0}$.
Therefore one root $\omega=\omega_{ci}$ pure oscillating mode while the other
root $\omega=\omega_{0}$ becomes the ion acoustic wave propagation mode i.e.,
$\omega_{0}\approx c_{s}k$ in the long wavelength limit $k\lambda_{D}\ll1$.
The quantum degeneracy effects appear only in the ion acoustic wave mode.

\item[(ii)] \textbf{Wave propagation in the perpendicular direction to the
magnetic field}: In case of exact perpendicular wave propagation i.e.,
$\theta=\pi/2$, the two symmetrical roots of Eq. (\ref{e60}) come out to be
$\omega=0$ (mode vanishes) and $\omega=\left(  \omega_{ci}^{2}+\omega_{0}%
^{2}\right)  ^{1/2}$. The ion cyclotron mode is modified in the quantum
effects in a magnetized plasma.

\item[(iii)] \textbf{Strongly magnetized ions case:} In case of a strongly
magnetized ions plasma i.e., $\omega_{ci}\gg \omega_{0}$, then from
Eq.(\ref{e60}) we have for the fast mode
\end{enumerate}

\begin{equation}
\omega^{2}=\omega_{ci}^{2}\left[  1+\frac{\omega_{0}^{2}}{\omega_{ci}^{2}}%
\sin^{2}\theta+\frac{\omega_{0}^{4}}{4\, \omega_{ci}^{4}}\sin^{2}%
(2\theta)+\mathcal{O}\left(  \left(  \frac{\omega_{0}}{\omega_{ci}}\right)
^{6}\right)  \right]  \,, \label{e63}%
\end{equation}
while the slow mode becomes
\begin{equation}
\omega^{2}=\omega_{0}^{2}\cos^{2}\theta \left[  1-\frac{\omega_{0}^{2}}%
{\omega_{ci}^{2}}\sin^{2}\theta+\mathcal{O}\left(  \left(  \frac{\omega_{0}%
}{\omega_{ci}}\right)  ^{4}\right)  \right]  \,. \label{e64}%
\end{equation}

\subsubsection{Kinetic Description}

The dispersion relation (\ref{e60}) obtained from fluid description must
agrees with the results of kinetic theory in the long wavelength limit.
Therefore, ionic and electronic susceptibilities obtained from kinetic theory
must be compared with Eqs. (\ref{e58}) and (\ref{e59}) obtained from fluid
dynamics. In order to find the ionic response, the particle distribution of
classical ions $f_{i}=f_{i}(\mathbf{r},\mathbf{v},t)$ satisfy the Vlasov's
equation as follows
\begin{equation}
\left[  \frac{\partial}{\partial t}+\mathbf{v}\cdot \nabla+\frac{e}{m_{i}%
}(-\nabla \phi+\mathbf{v}\times \mathbf{B}_{0})\cdot \frac{\partial}%
{\partial \mathbf{v}}\right]  f_{i}=0\,. \label{e65}%
\end{equation}
Due to quantum nature of electrons, the quantum Vlasov equation is satisfied
by the electronic Wigner quasi-distribution $f_{e}=f_{e}(\mathbf{r}%
,\mathbf{v},t)$,
\begin{align}
\frac{\partial f_{e}}{\partial t}  &  +\mathbf{v}\cdot \nabla f_{e}-\frac
{ie}{\hbar}\left(  \frac{m_{e}}{2\pi \hbar}\right)  ^{3}\times \int
d\mathbf{s}\,d\mathbf{v}^{\prime}\exp \left(  \frac{im_{e}(\mathbf{v}^{\prime
}-\mathbf{v})\cdot \mathbf{s}}{\hbar}\right)  \times \nonumber \\
&  \times \left[  \phi \left(  \mathbf{r}+\frac{\mathbf{s}}{2},t\right)
-\phi \left(  \mathbf{r}-\frac{\mathbf{s}}{2},t\right)  \right]  f_{e}%
(\mathbf{r},\mathbf{v}^{\prime},t)=0\,. \label{e66}%
\end{align}
All integrals run from $-\infty$ to $\infty$, unless otherwise stated. Also,
the magnetic force effects on electrons are ignored under the assumption of
large electron thermal and quantum (statistical and diffraction) effects.

The scalar potential is find out from Poisson's equation,
\begin{equation}
\left(  \frac{\partial^{2}}{\partial x^{2}}+\frac{\partial^{2}}{\partial
y^{2}}\right)  \phi=\frac{e}{\varepsilon_{0}}\left(  \int \!f_{e}%
\,d\mathbf{v}-\int \!f_{i}\,d\mathbf{v}\right)  \,, \label{e67}%
\end{equation}
where spatial variations are considered in the XY-plane.

The dispersion relation $1+\chi_{i}(\omega,\mathbf{k})+\chi_{e}(\omega
,\mathbf{k})=0$ for electrostatic waves in a magnetized quantum plasma is
easily obtained by assuming plane wave perturbations $\sim \exp[i(k_{x}%
x+k_{y}y-\omega t)]$ around isotropic distribution in velocities equilibria.
In case of cold ions and ignoring the Landau damping effects the ion
susceptibility obtained from kinetic theory (Stepanov 1959; Akhiezer et al.
1975) coincides well with its fluid description (\ref{e58}). However, under
the assumption of low frequency waves the static electronic response gives%

\begin{equation}
\chi_{e}(0,\mathbf{k})=\frac{e^{2}}{\varepsilon_{0}\hbar k^{2}}\int \!
\frac{d\mathbf{v}}{\mathbf{k}\cdot \mathbf{v}}\! \left[  F\left(
\mathbf{v}\!-\! \frac{\hbar \mathbf{k}}{2m_{e}}\right)  -F\left(
\mathbf{v}\!+\! \frac{\hbar \mathbf{k}}{2m_{e}}\right)  \right]  , \label{e68}%
\end{equation}
where the principal value of the integral is understood if necessary and where
the equilibrium electronic Wigner function is $f_{e}=F(\mathbf{v})$.

Now considering a Fermi-Dirac equilibrium for degenerate electrons,
\begin{equation}
F(\mathbf{v})=\frac{\mathcal{A}}{1+e^{\beta(m_{e}v^{2}/2-\mu_{(0)})}}\,,\quad
v=|\mathbf{v}|\,, \label{e69}%
\end{equation}
where normalization constant $\mathcal{A}$ assuring that $\int \!F(\mathbf{v}%
)\,d\mathbf{v}=n_{0}$.

Now using Eq. (\ref{e68}) in Eq. (\ref{e68}), the static electronic
susceptibility $\chi_{e}(0,\mathbf{k})$ can be expressed in the power series
of small quantum recoil $q=\sqrt{\beta/(2m_{e})}\, \hbar k/2$ for sufficiently
large electron temperature and long wavelength assumption i.e.,
\begin{align}
\chi_{e}(0,\mathbf{k})  &  =\frac{\beta m_{e}\omega_{pe}^{2}}{\sqrt{\pi}\,
\mathrm{Li}_{3/2}(-z)k^{2}}\left[  \Gamma \left(  \frac{1}{2}\right)
\mathrm{Li}_{1/2}(-z)+\Gamma \left(  -\frac{1}{2}\right)  \mathrm{Li}%
_{-1/2}(-z)\frac{q^{2}}{3}\right. \nonumber \\
&  \left.  +\Gamma \left(  -\frac{3}{2}\right)  \mathrm{Li}_{-3/2}%
(-z)\frac{q^{4}}{5}+\dots \right] \nonumber \\
&  =\frac{\beta m_{e}\omega_{pe}^{2}}{\sqrt{\pi}\, \mathrm{Li}_{3/2}(-z)k^{2}%
}\sum_{j=0}^{\infty}\Gamma \left(  \frac{1}{2}-j\right)  \times \mathrm{Li}%
_{1/2-j}(-z)\frac{q^{2j}}{2j+1}\,, \label{e70}%
\end{align}
where fugacity is $z=e^{\beta \mu_{(0)}}$. The above expression (\ref{e70}) is
exact, as long as the series converges and $\chi_{e}(0,\mathbf{k})$ has the
same expression as in Eq. (29) of (Melrose and Mushtaq 2010), where quantum
recoil correction of $\mathcal{O}(q^{2})$ was retained.

In order to have a comparison with fluid model results the expression
$\chi_{e}(0,\mathbf{k})$ described in Eq.(\ref{e59}) can be re-written in the
powers of small quantum recoil correction $q$ as follows,
\begin{align}
\chi_{e}(0,\mathbf{k})  &  =\frac{\beta m_{e}\omega_{pe}^{2}\mathrm{Li}%
_{1/2}(-z)}{\, \mathrm{Li}_{3/2}(-z)k^{2}}\left(  1+\frac{2q^{2}}{3}%
\frac{\mathrm{Li}_{-1/2}(-z)}{\mathrm{Li}_{1/2}(-z)}\right)  ^{-1}\nonumber \\
&  =\frac{\beta m_{e}\omega_{pe}^{2}}{\, \mathrm{Li}_{3/2}(-z)k^{2}}\left[
\mathrm{Li}_{1/2}(-z)-\frac{2q^{2}}{3}\mathrm{Li}_{-1/2}(-z)+\mathcal{O}%
(q^{4})\, \right]  \text{ ,} \label{e71}%
\end{align}
which is the same as described in Eq. (\ref{e70}) in the long wavelength
limit, where $\Gamma(1/2)=\sqrt{\pi},\, \, \Gamma(-1/2)=-2\sqrt{\pi}$.
It\ also justifies the inclusion numerical parameter $\alpha$ in the quantum
force of Eq. (\ref{e55}) in the magnetized dense plasma case to compare the
results of kinetic theory with the fluid description of 3D Fermi-Dirac
distributed electrons.

\subsection{Zakharov-Kuznetsov (ZK) Equation for Ion-acoustic Waves in a
Magnetized Dense Plasma}

In this section, we will derive ZK equation for the two dimensional
propagating ion-acoustic soliton in a magnetized quantum plasma with arbitrary
degeneracy of electrons. The set of dynamic equations (\ref{e51})-(\ref{e56})
can be written in normalized or dimensionless form as follows,%

\begin{equation}
\frac{\partial \tilde{n}_{i}}{\partial \tilde{t}}+\frac{\partial}{\partial
\tilde{x}}(\tilde{n}_{i}\tilde{u}_{ix})+\frac{\partial}{\partial \tilde{y}%
}(\tilde{n}_{i}\tilde{u}_{iy})=0\,, \label{e72}%
\end{equation}%
\begin{equation}
\frac{\partial \tilde{u}_{ix}}{\partial \tilde{t}}+\left(  \tilde{u}_{ix}%
\frac{\partial}{\partial \tilde{x}}+\tilde{u}_{iy}\frac{\partial}%
{\partial \tilde{y}}\right)  \tilde{u}_{ix}=-\frac{\partial \tilde{\phi}%
}{\partial \tilde{x}}\,, \label{e73}%
\end{equation}%
\begin{equation}
\frac{\partial \tilde{u}_{iy}}{\partial \tilde{t}}+\left(  \tilde{u}_{ix}%
\frac{\partial}{\partial \tilde{x}}+\tilde{u}_{iy}\frac{\partial}%
{\partial \tilde{y}}\right)  \tilde{u}_{iy}=-\frac{\partial \tilde{\phi}%
}{\partial \tilde{y}}+\Omega \tilde{u}_{iz}\,, \label{e74}%
\end{equation}%
\begin{equation}
\frac{\partial \tilde{u}_{iz}}{\partial \tilde{t}}+\left(  \tilde{u}_{ix}%
\frac{\partial}{\partial \tilde{x}}+\tilde{u}_{iy}\frac{\partial}%
{\partial \tilde{y}}\right)  \tilde{u}_{iz}=-\Omega \tilde{u}_{iy}\,,
\label{e750}%
\end{equation}%
\begin{equation}
0=\tilde{\nabla}\tilde{\phi}-\frac{\mathrm{Li}_{1/2}(-e^{\beta \mu_{(0)}}%
)}{\mathrm{Li}_{1/2}(-e^{\beta \mu})}\tilde{\nabla}\tilde{n}_{e}+\frac{H^{2}%
}{2}\tilde{\nabla}\left[  \frac{1}{\sqrt{\tilde{n}_{e}}}\left(  \frac
{\partial^{2}}{\partial \tilde{x}^{2}}+\frac{\partial^{2}}{\partial \tilde
{y}^{2}}\right)  \sqrt{\tilde{n}_{e}}\right]  \label{e76}%
\end{equation}%
\begin{equation}
\left(  \frac{\partial^{2}}{\partial \tilde{x}^{2}}+\frac{\partial^{2}%
}{\partial \tilde{y}^{2}}\right)  \tilde{\phi}=\tilde{n}_{e}-\tilde{n}_{i},
\label{e77}%
\end{equation}
where $\Omega=\omega_{ci}/\omega_{pi}$ and $\tilde{\nabla}=(\partial
/\partial \tilde{x},\partial/\partial \tilde{y},0)$ have been defined. The
expression of normalized electron density for arbitrary degenerate electrons
has already been described in Eq. (\ref{e34}). The normalization of dependent
variables $(\tilde{u}_{ix},\tilde{u}_{iy},\tilde{u}_{iz})=(u_{ix}%
,u_{iy},u_{iz})/c_{s}$, $\tilde{\phi}=e\phi/(m_{i}c_{s}^{2})$, $\tilde{n}%
_{j}=n_{j}/n_{0}$, where $j=e,i$, while for independent variables $(\tilde
{x},\tilde{y})=(x,y)/\lambda_{D}$, $\tilde{t}=\omega_{_{pi}}t$ has been
defined. For simplicity, the tilde sign will be omitted from normalized
quantities in further calculations.

In order to find a nonlinear evolution equation for obliquely propagating
ion-acoustic wave in a dense magnetized plasma, the stretching of the
independent variables ($x,$ $y,$ $t$) is defined under the assumption of
strong magnetization as follows (Kourakis et al. 2009; Mace and Hellberg 2001;
Infeld 1985; Laedke and Spatschek 1982),
\begin{equation}
X=\varepsilon^{1/2}(x-V_{0}t)\,,\quad Y=\varepsilon^{1/2}y\,,\quad
\tau=\varepsilon^{3/2}t\,,
\end{equation}
where $\varepsilon$ is a small expansion parameter and $V_{0}$ is the phase
velocity of the wave, to be determined later on. The perturbed quantities can
be expanded in powers of $\varepsilon$ as follows,
\begin{align}
n_{j}  &  =1+\varepsilon n_{j1}+\varepsilon^{2}n_{j2}+...\,,\quad j=e,i\,,\\
u_{ix}  &  =\varepsilon u_{x1}+\varepsilon^{2}u_{x2}+\varepsilon^{3}%
u_{x3}...\,,\\
u_{i\perp}  &  =\varepsilon^{3/2}u_{\perp1}+\varepsilon^{2}u_{\perp
2}+\varepsilon^{5/2}u_{\perp3}...\,,\quad \perp=y,z\,,\\
\phi &  =\varepsilon \phi_{1}+\varepsilon^{2}\phi_{2}+...\,,\\
\mu &  =\mu_{(0)}+\varepsilon \mu_{1}+\varepsilon^{2}\mu_{2}+... \label{e79}%
\end{align}
In the above perturbation scheme, the ion fluid velocity components ($u_{iy}$,
$u_{iz}$) in the perpendicular direction of the external magnetic field are
taken to be of higher order perturbations in comparison with the ion parallel
fluid velocity component $u_{ix}$ because in the presence of magnetic field
the anisotropy of fluid velocities occurs and ion gyro-motion becomes a higher
order effect.

Now collecting the lowest $\varepsilon$ order terms from the set of dynamic
equations (\ref{e72})-(\ref{e77}) and on solving \ them together, we obtain
$V_{0}=\pm1$, which is the normalized phase velocity of the ion-acoustic waves
in a magnetized dense plasma. In further calculations, we set $V_{0}=1$
without loss of generality.

Collecting the next higher $\varepsilon$ order terms from the set of dynamic
equations (\ref{e72})-(\ref{e77}),
it is finally possible to write the ZK equation for obliquely propagating
ion-acoustic wave in a magnetized quantum plasma in terms of $\phi_{1}%
\equiv \varphi$,
\begin{equation}
\frac{\partial \varphi}{\partial \tau}+A\varphi \frac{\partial \varphi}{\partial
X}+\frac{\partial}{\partial X}\left(  B\frac{\partial^{2}\varphi}{\partial
X^{2}}+C\frac{\partial^{2}\varphi}{\partial Y^{2}}\right)  =0\,, \label{e85}%
\end{equation}
where nonlinearity coefficient $A$ and the dispersion coefficients in the
parallel $(B)$ and perpendicular $(C)$ directions of the external magnetic
field, respectively, are defined as
\begin{align}
A  &  =\frac{1}{2}\left(  3-\alpha \right)  \,,\label{e86}\\
B  &  =\frac{1}{2}\left(  1-\frac{H^{2}}{4}\right)  \,,\label{e87}\\
C  &  =\frac{1}{2}\left(  1+\frac{1}{\Omega^{2}}-\frac{H^{2}}{4}\right)  \,.
\label{e88}%
\end{align}
The limiting cases of the dilute and dense plasmas are discussed here i.e., in
a low fugacity $e^{\beta \mu_{(0)}}<<1$ or dilute plasma case, the nonlinearity
and dispersion coefficients come out to be $A=1,$ $B=1/2$ and $C=(1/2)\left(
1+1/\Omega^{2}\right)  $, which are the same as in Refs.(Zakharov and
Kuznetsov 1974; Infeld 1985; Laedke and Spatschek 1982) for the ZK equation of
nonlinear ion-acoustic waves in a classical magnetized plasma. However, for
the dense magnetized plasma case i.e., $e^{\beta \mu_{(0)}}>>1$, we will have
will have $A=4/3$ , while $H=(1/2)\, \hbar \omega_{pe}/\varepsilon_{F}$ appears
in the dispersive coefficients $B,C$. The ZK equation (\ref{e85}) for fully
degenerate magnetized plasma case does not matches exactly with the results of
(Sabry et al. 2008; Moslem et al. 2007; Khan and Masood 2008) but our results
come out to be same in the limiting case of classical electron-ion magnetized
plasma already published in the literature.

The traveling wave soliton solution of the ZK equation (\ref{e85}) for
obliquely propagating ion-acoustic waves in a magnetized dense plasmas is%

\begin{equation}
\varphi=\varphi_{0}\, \mathrm{sech}^{2}(\eta/W)\,, \label{e89}%
\end{equation}
where $\varphi_{0}=3u_{0}/(Al_{x})$ is the height and $W=\sqrt{4l_{x}%
(l_{x}^{2}B+l_{y}^{2}C)/u_{0}}$ is the width of the soliton. The transformed
coordinate $\eta$ in the co-moving frame is defined as $\eta=l_{x}%
X+l_{y}Y-u_{0}\tau$, where $u_{0}$ is the speed of the soliton and $l_{x}>0$
and $l_{y}$ are direction cosines, so that $l_{x}^{2}+l_{y}^{2}=1$. The
formation of potential hump (bright) and dip (dark) structure of soliton
depends on the the sign of $\varphi_{0}$. The decaying boundary conditions
have been used at infinity for soliton structure. The dispersion effects of
both charge separation and ion Larmor radius balances with the nonlinearity in
the system to form a soliton structure of ZK equation.

The soliton solution in the laboratory frame can be written as
\begin{equation}
\varphi=\frac{3\, \delta V}{A}\, \mathrm{sech}^{2}\left[  \frac{1}{2}\left(
\frac{\delta V}{l_{x}^{2}B+l_{y}^{2}C}\right)  ^{1/2}\left \{  l_{x}\left(
x-(V_{0}+\delta V)\,t\right)  +l_{y}y\right \}  \right]  \label{e90}%
\end{equation}
where $\delta V=\varepsilon u_{0}/l_{x}$ is defined and $V_{0}+\delta V$
corresponds to the velocity at which travels the intersection between a plane
of constant phase and a field line, down the same field line (Mace and
Hellberg 2001). Also, the width $L$ of the soliton in the laboratory frame is
given by
\begin{equation}
L=2\left(  \frac{l_{x}^{2}B+l_{y}^{2}C}{\delta V}\right)  ^{1/2}=\sqrt
{\frac{2}{\delta V}}\left(  1-\frac{H^{2}}{4}+\frac{l_{y}^{2}}{\Omega^{2}%
}\right)  ^{1/2} \label{e91}%
\end{equation}
It is evident from above expression that the soliton width $L$ remains real
and positive in the classical plasma limit i.e., $H=0$ for $\delta V>0$, which
means hump (bright) soliton structure moving with supersonic speed are formed.
However, in case of quantum plasma there seems possibility of formation of
both bright and dark solitons depending on the condition of quantum parameter
$H$. Theoretically speaking there is always a possibility of dark soliton
formation under condition $H^{2}/4>1+l_{y}^{2}/\Omega^{2}$ and $\delta V<0$
moving with subsonic speed. The amplitude of the soliton depends on
nonlinearity coefficient $A$, which is independent of quantum diffraction
parameter and depends only on quantum degeneracy. In case of more degenerate
system the degeneracy parameter $\alpha$ has smaller values and therefore the
numerical value of nonlinearity coefficient $A$ is increased which results in
the decrease in soliton amplitude.

Now in case of unmagnetized quantum plasma term $\sim1/\Omega^{2}$ does not
appear in the ZK equation (\ref{e85}) and perpendicular dispersive coefficient
$C$ also disappears to give a KdV equation. Also, at $H=2$ the corresponding
KdV equation collapses to the Burger's equation, producing an ion-acoustic
shock wave structure instead of a soliton (Haas et al. 2003). In a magnetized
quantum plasma another possibility happens for ZK equation (\ref{e85}) in
which $C=0$ if $1+1/\Omega^{2}=H^{2}/4$, we have
\begin{equation}
\frac{\partial \varphi}{\partial \tau}+A\varphi \frac{\partial \varphi}{\partial
X}-\frac{1}{2\Omega^{2}}\frac{\partial^{3}\varphi}{\partial X^{3}}=0\,,
\label{e92}%
\end{equation}
which transforms to the KdV equation in its standard form only if
$\varphi \rightarrow-\varphi,X\rightarrow-X,\tau \rightarrow \tau$ and becomes
completely integrable.

{Finally, if $B=0$, which happens only if $1-H^{2}/4=0$ and dispersive
coefficient in the perpendicular direction becomes$C=1/2\Omega^{2}$, so that
Eq. (\ref{e85}) can be re written as
\begin{equation}
\frac{\partial \varphi}{\partial \tau}+A\varphi \frac{\partial \varphi}{\partial
X}+\frac{1}{2\Omega^{2}}\frac{\partial}{\partial X}\frac{\partial^{2}\varphi
}{\partial Y^{2}}=0, \label{e93}%
\end{equation}
This is a KdV-like equation having dispersive effects in the direction
perpendicular to the magnetic field.}


\begin{figure}[tbh]
\begin{center}
\includegraphics[width=8cm]{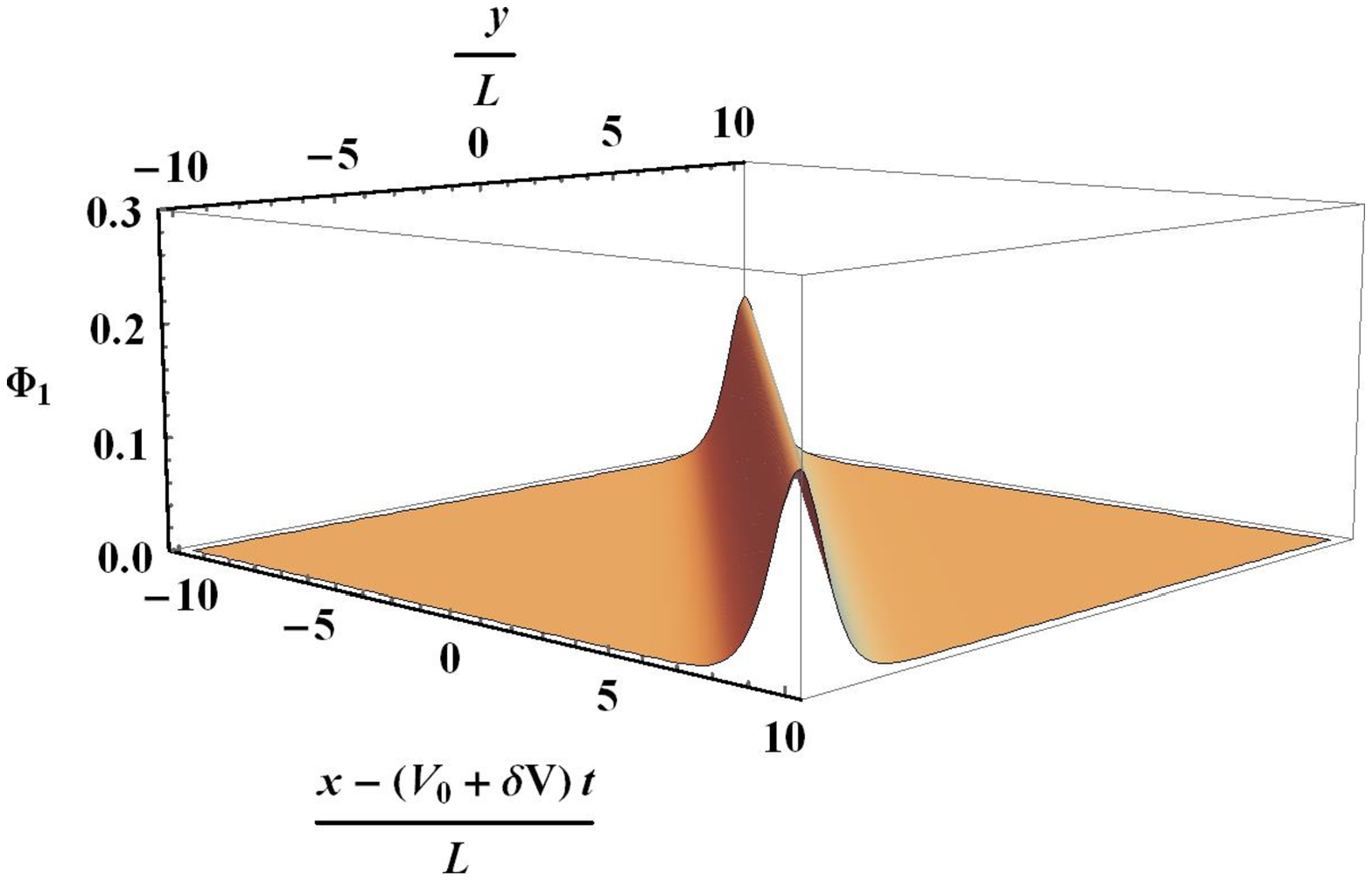}
\end{center}
\caption{The ion-acoustic hump soliton structure (\ref{e90}) in a magnetized
quantum plasma (moving with supersonic speed in the laboratory frame) is shown
for normalized parameters: $\alpha=0.5,H^{2}=0.1,$ $\delta V=0.1$,
$l_{x}=l_{y}=$ $\sqrt{2}/2$ and $g=0.22$ $(<1)$. }%
\label{figure4}%
\end{figure}

In order to investigate the valid range of the coefficients $A,$ $B$ and $C$
defined in Eqs.(\ref{e86}--\ref{e88}), respectively, for the existence of
ion-acoustic soliton in a quantum magnetized plasma it is observed that
$1<A<4/3$ for $1/3<\alpha<1$. Also, $H^{2}$ from Eq. (\ref{e330}) in principle
can have large values due to which dispersive coefficients $B$ and $C$ do not
remain positive definite in dense plasmas. The strong coupling effects
associated with large values of $H^{2}$ can have its significant impact on the
propagation of quantum ion-acoustic soliton, which could be addressed in a
separate extended theory. Therefore, the existence or nonexistence of pure
quantum ion-acoustic soliton in such a generalized study needs more precise
statements. However, still some significant values of $H^{2}$ are physically
acceptable in the present study. The two dimensional ion-acoustic potential
hump (bright) soliton moving with supersonic speed in the laboratory frame is
shown in Fig. 4 for plasma parameters for solid density i.e., $n_{0}%
=5\times10^{30}\, \mathrm{m}^{-3},T=T_{F}=1.24\times10^{6}\, \mathrm{K}$ and
$B_{0}=10^{3}\, \mathrm{T}$ for which $\alpha=0.5,$ $H^{2}=0.1$.

\section{\textbf{Model for Magnetosonic Waves with Arbitrary Degeneracy of
Electrons in Quantum Plasmas}}

In this section, we will investigate the magnetosonic waves in a magnetized
quantum plasma with arbitrary degeneracy of electrons. The two fluid model is
applied in which length scales remains less than the electron skin depth. The
ions are assumed to be classical due to their large mass, while degenerate
electrons contains Fermi pressure and quantum force due to wave like nature.
This section is mainly a review from the material in (Haas and Mahmood 2018).
The quantum fluid equations for magnetosonic waves in a dense plasma are
described as follow.

The ion fluid continuity and momentum equations are given by%
\begin{equation}
\frac{\partial n_{i}}{\partial t}+\mathbf{\nabla}\cdot \left(  n_{i}%
\mathbf{v}_{i}\right)  =0, \label{e100}%
\end{equation}%
\begin{equation}
\frac{\partial \mathbf{v}_{i}}{\partial t}+\left(  \mathbf{v}_{i}%
.\mathbf{\nabla}\right)  \mathbf{v}_{i}=\frac{e}{m_{i}}\left(  \mathbf{E}%
+\mathbf{v}_{i}\times \mathbf{B}\right)  , \label{e101}%
\end{equation}
where the ion fluid density and velocity are represented by $n_{i}$ and
$\mathbf{v}_{i}$ respectively, while the ionic number density $\mathbf{E}$,
$\mathbf{B}$ stands for the electromagnetic field, $m_{i}$ is the ions mass,
and $-e$ is the electronic charge.

The continuity and momentum equations for electron quantum fluid is written as%
\begin{equation}
\frac{\partial n_{e}}{\partial t}+\mathbf{\nabla}\cdot \left(  n_{e}%
\mathbf{v}_{e}\right)  =0, \label{e102}%
\end{equation}%
\begin{equation}
\frac{\partial \mathbf{v}_{e}}{\partial t}+\left(  \mathbf{v}_{e}%
.\mathbf{\nabla}\right)  \mathbf{v}_{e}=-\frac{e}{m_{e}}\left(  \mathbf{E}%
+\mathbf{v}_{e}\times \mathbf{B}\right)  -\frac{1}{n_{e}m_{e}}\mathbf{\nabla
}p+\left(  \frac{\alpha}{3}\right)  \frac{\hslash^{2}}{2m_{e}^{2}%
}\mathbf{\nabla}\left(  \frac{1}{\sqrt{n_{e}}}\nabla^{2}\sqrt{n_{e}}\right)  ,
\label{e103}%
\end{equation}
where electron number density and fluid velocity are represented by $n_{e}$
and $\mathbf{v}_{e}$ , $m_{e}$ is the electron mass, $\hbar$ is the reduced
Planck constant, $p$ is the electronic fluid pressure. The last term is the
Bohm potential due to wave like nature of electrons and $\alpha$ is a
numerical factor, which is set for comparison of results from fluid and
kinetic theory in a 3D Fermi-Dirac distributed electrons plasma. The obtained
expression of $\alpha$ in terms of fugacity $e^{\beta \mu_{(0)}}$ for low
frequency waves is described in Eq.(\ref{e293}) and its numerical values in
limiting cases are i.e., for classical plasma ($\alpha \approx1$) and for fully
degenerate case ($\alpha \approx1/3$) .

In order to derive the equation of state for electron pressure, again we
consider a local quasi-equilibrium Fermi-Dirac distribution function defined
in Eq. (\ref{e9}). The relation between electron density $n_{e}$ and chemical
potential $\mu$ is defined in Eq. (\ref{e8}), while the relation between
equilibrium electron density and equilibrium chemical potential is defined in
(\ref{e801}). Now using Eqs.(\ref{e8}) and (\ref{e801}), which gives a
relation of electron density $n_{e}$ containing $n_{0}$, $\mu$ and $\mu_{(0)}%
$, which is used in Eq. (\ref{e9}) before inserting equation of state for
electron pressure in Eq. (\ref{e103}).

The Faraday's and Ampere's laws are written as
\begin{equation}
\mathbf{\nabla}\times \mathbf{E}=-\frac{\partial \mathbf{B}}{\partial t},
\label{e104}%
\end{equation}%
\begin{equation}
\mathbf{\nabla}\times \mathbf{B}=\mu_{0}\mathbf{J+}\frac{1}{c^{2}}%
\frac{\partial \mathbf{E}}{\partial t}, \label{e105}%
\end{equation}
where $\mu_{0}$ is the free space permeability and $c$ the speed of light. The
current density is defined as
\begin{equation}
\mathbf{J}=e(n_{i}\mathbf{v}_{i}-n_{e}\mathbf{v}_{e}). \label{e106}%
\end{equation}
The charge neutrality condition $n_{i}\approx n_{e}$ is assumed and in
equilibrium we have $n_{e,i}=n_{0}$, $\mathbf{v}_{e,i}=0$, $\mathbf{E}=0$ and
$\mathbf{B}=B_{0}\hat{z}$, a uniform magnetic field. The magnetosonic waves
are low frequency waves i.e., $\omega<<\Omega_{i}$, where $\Omega_{i}%
=eB_{0}/m_{i}$ is the ion gyro-frequency and also the displacement current in
Eq. (\ref{e105}) will be ignored under the assumption that phase speed of the
wave is much less than the speed of light.

\subsection{Linear Analysis}

In order to find the dispersion relation of the magnetosonic waves in a
magnetized quantum plasma with arbitrary degeneracy of electron, we have to
linearized the set dynamic Eqs.(\ref{e100})-(\ref{e106}). The external
magnetic field is assumed to be directed along $z$-axis i.e., $\mathbf{B}%
_{0}=B_{0}\hat{z}$ and wave propagation is taken along x-axis i.e.,
$\mathbf{\nabla=(\partial}_{x},0,0)$. The electric field lies in the XY-plane
i.e., $\mathbf{E}=E_{x}\hat{x}+E_{y}\hat{y}$ and magnetic field is
$\mathbf{B}=(B_{0}+B_{z})\hat{z}$. Now assuming the sinusoidal perturbations
of the form $\exp \left(  ikx-i\omega t\right)  $, the obtained dispersion
relation of magnetosonic waves in dense magnetized plasma is given by%

\begin{equation}
\omega^{2}=\left(  c_{s}^{2}+\frac{\alpha}{12}\frac{\hslash^{2}}{m_{i}m_{e}%
}k^{2}\right)  k^{2}+\frac{k^{2}v_{A}^{2}}{1+k^{2}\lambda_{e}^{2}},
\label{e107}%
\end{equation}
where\ $c_{s}$ is the ion-acoustic speed defined in Eq. (\ref{e23}) and
$v_{A}=B_{0}/\sqrt{\mu_{0}m_{i}n_{0}}$ is the Alfv\'{e}n speed. The electron
skin depth $\lambda_{e}=\sqrt{c/\omega_{pe}}$ is defined, where $\omega
_{pe}=\sqrt{n_{0}e^{2}/(m_{e}\varepsilon_{0})}$ is the electrons plasma
frequency with $\varepsilon_{0}$ being the free space permittivity.

In case of short wavelength i.e., $k^{2}\lambda_{e}^{2}>>1$, the magnetosonic
dispersion relation (\ref{e107}) gives,%
\begin{equation}
\omega^{2}=\left(  c_{s}^{2}+\frac{\alpha}{12}\frac{\hslash^{2}}{m_{i}m_{e}%
}k^{2}\right)  k^{2}+\Omega_{i}\Omega_{e}, \label{e108}%
\end{equation}
which is the lower hybrid frequency mode in a magnetized plasma modified in
the presence quantum effects. Here $\Omega_{j}=eB_{0}/m_{j}$ (here $j=i,e$) is
the gyrofrequency of the plasma species.

It can be seen from dispersion relation (\ref{e107}) that wave dispersion
effects appears due to finite electron skin depth $\lambda_{e}$ and inclusion
of Bohm potential effects in the model. The phase velocity of the magnetosonic
waves in the absence of Bohm term and in the long wavelength ( $k^{2}%
\lambda_{e}^{2}<<1$) is given by%

\begin{equation}
\frac{\omega}{k}=\sqrt{c_{s}^{2}+v_{A}^{2}}, \label{e141}%
\end{equation}
The limiting case analysis of non-degenerate and fully degenerate magnetized
plasmas of obtained dispersion relation of magnetosonic waves in a quantum
plasma with arbitrary degeneracy of electrons is done below:

In case of dilute plasma case with a small fugacity $e^{\beta \mu_{(0)}}\ll1$
and ignoring the Bohm potential, then dispersion relation (\ref{e107}) gives,
\begin{equation}
\omega^{2}=c_{s}^{2}k^{2}+\frac{k^{2}v_{A}^{2}}{1+k^{2}\lambda_{e}^{2}},
\label{e109}%
\end{equation}
which is the same as Eq. (16) in Ref. (Ur-Rehman et al. 2017), while
ion-acoustic speed is $c_{s}=\sqrt{\kappa_{B}T/m_{i}}$ defined for a classical plasma.

However, for fully degenerate plasma case i.e., fugacity $e^{\beta \mu_{(0)}%
}\ll1$, the dispersion relation (\ref{e107}) for magnetosonic waves is written as%

\begin{equation}
\omega^{2}=\left(  c_{s}^{2}+\frac{\hslash^{2}}{36m_{i}m_{e}}k^{2}\right)
k^{2}+\frac{k^{2}v_{A}^{2}}{1+k^{2}\lambda_{e}^{2}}, \label{e110}%
\end{equation}
where $c_{s}=\sqrt{2\varepsilon_{F}/(3m_{i})}$ is the ion-acoustic speed in a
quantum plasma, while $\varepsilon_{F}=\kappa_{B}T_{F}=[\hbar^{2}%
/(2m_{e})]\,(3\pi^{2}n_{0})^{2/3}$ is the Fermi energy of electrons and in
fully degenerate plasmas $\mu_{(0)}=\varepsilon_{F}$.

\subsection{Derivation of a KdV equation for magnetosonic solitons in dense
plasmas with arbitrary degeneracy of electrons}

In order to derive the KdV equation for magnetosonic solitons in dense plasma
with arbitrary degeneracy of electrons, the set of dynamic equations
(\ref{e100})-(\ref{e106}) can be written in normalized form as follows:

The ion continuity equation is written as%
\begin{equation}
\frac{\partial \tilde{n}_{i}}{\partial \tilde{t}}+\frac{\partial}{\partial
\tilde{x}}\left(  \tilde{n}_{i}\tilde{v}_{ix}\right)  =0. \label{e111}%
\end{equation}
The ion momentum equation along $x$ and $y$ axis are described by%
\begin{equation}
\frac{\partial \tilde{v}_{ix}}{\partial \tilde{t}}+\tilde{v}_{ix}\frac{\partial
}{\partial \tilde{x}}\tilde{v}_{ix}=\tilde{E}_{x}+\Omega \text{ }\tilde{v}%
_{iy}\tilde{B}, \label{e112}%
\end{equation}%
\begin{equation}
\frac{\partial \tilde{v}_{iy}}{\partial \tilde{t}}+\tilde{v}_{ix}\frac{\partial
}{\partial \tilde{x}}\tilde{v}_{iy}=\tilde{E}_{y}-\Omega \text{ }\tilde{v}%
_{ix}\tilde{B}. \label{e113}%
\end{equation}
The continuity equation of electron is written as
\begin{equation}
\frac{\partial \tilde{n}_{e}}{\partial \tilde{t}}+\frac{\partial}{\partial
\tilde{x}}\left(  \tilde{n}_{e}\tilde{v}_{ex}\right)  =0, \label{e114}%
\end{equation}
The $x$ and $y$ components of the electron fluid momentum equation are given
by%
\begin{align}
&  \left.  \frac{\partial \tilde{v}_{ex}}{\partial \tilde{t}}+\tilde{v}%
_{ex}\frac{\partial}{\partial \tilde{x}}\tilde{v}_{ex}=-\delta \text{ }\tilde
{E}_{x}-\delta \text{ }\Omega \text{ }\tilde{v}_{ey}\tilde{B}\right.
-\delta \text{ }\frac{\mathrm{Li}_{1/2}(-e^{\beta \mu_{(0)}})}{\mathrm{Li}%
_{1/2}(-e^{\beta \mu})}\frac{\partial}{\partial \tilde{x}}\tilde{n}%
_{e}\nonumber \\
&  \left.  +\frac{1}{2}\delta \text{ }H^{2}\text{ }\frac{\partial}%
{\partial \tilde{x}}\left(  \frac{1}{\sqrt{\tilde{n}_{e}}}\frac{\partial^{2}%
}{\partial \tilde{x}^{2}}\sqrt{\tilde{n}_{e}}\right)  ,\right.  \label{e115}%
\end{align}%
\begin{equation}
\frac{\partial \tilde{v}_{ey}}{\partial \tilde{t}}+\tilde{v}_{ex}\frac{\partial
}{\partial \tilde{x}}\tilde{v}_{ey}=-\delta \text{ }\tilde{E}_{y}+\delta \text{
}\Omega \text{ }\tilde{v}_{ex}\tilde{B}. \label{e116}%
\end{equation}
The $z$ component of the Faraday's law yields%
\begin{equation}
\frac{\partial \tilde{E}_{y}}{\partial \tilde{x}}=-\Omega \text{ }\frac
{\partial \tilde{B}}{\partial \tilde{t}}. \label{e117}%
\end{equation}
The $x$ and $y$ components of Ampere's law are written as%
\begin{equation}
0=\tilde{n}_{i}\tilde{v}_{ix}-\tilde{n}_{e}\tilde{v}_{ex}, \label{e118}%
\end{equation}%
\begin{equation}
\Omega \text{ }\frac{\partial \tilde{B}}{\partial \tilde{x}}=\frac{c_{s}^{2}%
}{c^{2}}\left(  \tilde{n}_{e}\tilde{v}_{ey}-\tilde{n}_{i}\tilde{v}%
_{iy}\right)  . \label{e119}%
\end{equation}
The normalization of electron and ion fluid density and fluid velocities are
defined as $\tilde{n}_{e,i}=n_{e,i}/n_{0}$ and $\mathbf{\tilde{v}}%
_{e,i}=\mathbf{v}_{e,i}/c_{s}$, respectively. Also, the normalization of
space, time, electric and magnetic fields i.e., $\tilde{x}=\omega
_{pi}\,x\mathbf{/}c_{s}$,$\  \tilde{t}=\omega_{pi}t$,$\  \tilde{\mathbf{E}%
}\mathbf{=}e\mathbf{E}\mathbf{/}\left(  m_{i}c_{s}\omega_{pi}\right)  $ and
$\tilde{\mathbf{B}}=B\hat{z}\mathbf{/}B_{0}$ are defined. The dimensionless
quantum diffraction parameter $H$ is defined in Eq. (\ref{e330}) and
$\delta=m_{i}/m_{e}$ is the ion-electron mass ratio defined in Eq.
(\ref{e115}).\ The ratio of ion-cyclotron to ion plasma frequencies is defined
as $\Omega=\Omega_{i}/\omega_{pi}$, where ion plasma frequency is $\omega
_{pi}=\sqrt{e^{2}n_{0}/(m_{i}\varepsilon_{0})}$. The normalized electron
density ($\tilde{n}_{e}$) with chemical potential ($\mu$) are related through
relation described in Eq. (\ref{e34}). In further calculations, the tilde sign
on normalized variables will not be used for simplicity.

The well known reductive perturbation method is employed (Hussain and Mahmood
2017) to find the KdV equation and stretching spatial and temporal independent
variables is defined as follows,%
\[
\xi=\epsilon^{1/2}(x-v_{0}t)\text{, \  \  \  \  \  \  \  \  \  \  \  \ }\tau
=\epsilon^{3/2}t.
\]
Here $\epsilon$ is a small expansion parameter which characterizes the
strength of nonlinearity and $v_{0}$ is the normalized phase velocity of the
wave to be determined later on.

The expansion of dynamic variables expanded in terms of the smallness
parameter $\epsilon$ is defined below (where $j=e,i$),%
\begin{align}
n_{j}  &  =1+\epsilon n_{j1}+\epsilon^{2}n_{j2}+...\nonumber \\
v_{jx}  &  =\epsilon v_{jx1}+\epsilon^{2}v_{jx2}+...\nonumber \\
v_{jy}  &  =\epsilon^{3/2}v_{jy1}+\epsilon^{5/2}v_{jy2}+...\nonumber \\
E_{x}  &  =\epsilon^{3/2}E_{x1}+\epsilon^{5/2}E_{x2}+...\nonumber \\
E_{y}  &  =\epsilon E_{y1}+\epsilon^{2}E_{y2}+...\nonumber \\
B  &  =1+\epsilon B_{z1}+\epsilon^{2}B_{z2}+....\nonumber \\
\mu &  =\mu_{(0)}+\epsilon \mu_{1}+\epsilon^{2}\mu_{2}+... \label{e121}%
\end{align}
Now applying the perturbation scheme in Eqs. (\ref{e112}) to (\ref{e119}) and
collecting the lowest order terms, which gives normalized phase velocity\ of
the wave as follows:
\begin{equation}
v_{0}=\pm \sqrt{1+\frac{v_{A}^{2}}{c_{s}^{2}}}, \label{e122}%
\end{equation}
which is the same phase velocity of the wave as already derived in Eq.
(\ref{e141}) for a dispersionless wave. The positive sign of $v_{0}$ will be
retained in further calculations without any loss of generality.

Now collecting the next higher order ($\varepsilon$) terms from the set of
dynamic equations and on solving them we obtain,%
\begin{equation}
\frac{f_{2}}{v_{0}}+f_{3}+\frac{f_{5}}{\delta}-\frac{\Omega c^{2}}{c_{s}^{2}%
}f_{9}-\frac{\Omega c^{2}}{v_{0}c_{s}^{2}}f_{7}+\frac{1}{\delta}\left(
\frac{\Omega c^{2}}{v_{0}c_{s}^{2}}\right)  \frac{\partial f_{6}}{\partial \xi
}=0. \label{e123}%
\end{equation}
Equation (\ref{e122}) has been used in the derivation of Eq. (\ref{e123}) and
the expression of $f_{1}$ to $f_{9}$ are described in the Appendix.

In order to express the first order perturbed quantities $v_{ix1}$, $n_{i1}$,
$n_{e1}$, $B_{z1}$, $E_{x1}$, $E_{y1}$, $v_{iy1}$ and $v_{ey1}$\ in terms of
one variable $v_{ex1}$, we have%
\begin{equation}
E_{y1}=\Omega v_{ex1}, \label{e124}%
\end{equation}%
\begin{equation}
v_{ix1}=v_{ex1}, \label{e125}%
\end{equation}%
\begin{equation}
n_{i1}=n_{e1}=\frac{v_{ex1}}{v_{0}}, \label{e126}%
\end{equation}%
\begin{equation}
B_{z1}=\frac{v_{ex1}}{v_{0}}, \label{e127}%
\end{equation}%
\begin{equation}
E_{x1}=-\frac{1}{v_{0}\delta}\left(  v_{0}^{2}\, \delta+1\right)
\frac{\partial v_{ex1}}{\partial \xi}\approx-v_{0}\frac{\partial v_{ex1}%
}{\partial \xi}, \label{e128}%
\end{equation}%
\begin{equation}
v_{iy1}=\frac{1}{\Omega v_{0}\delta}\frac{\partial v_{ex1}}{\partial \xi},
\label{e129}%
\end{equation}%
\begin{equation}
v_{ey1}=\frac{1}{\Omega v_{0}}\left(  v_{0}^{2}+1\right)  \frac{\partial
v_{ex1}}{\partial \xi}, \label{e130}%
\end{equation}
where the relation $v_{ey1}=-\delta \,v_{iy1}$ between electron and ion
momentum along $y$ direction has also been used \ to derive the above
relations. Also, the approximations $\delta \gg1,v_{0}>1$ has been used to
obtain relation (\ref{e128}).

Using relations (\ref{e124})-(\ref{e130}) in Eq. (\ref{e123}) and after some
simplification, the KdV equation is obtained for magnetosonic waves in a
quantum plasma in terms $v_{ex1}$ is given by%
\begin{equation}
\frac{\partial v_{ex1}}{\partial \tau}+aa\text{ }v_{ex1}\frac{\partial v_{ex1}%
}{\partial \xi}+bb\text{ }\frac{\partial^{3}v_{ex1}}{\partial \xi^{3}}=0,
\label{e131}%
\end{equation}
where the nonlinearity and dispersion coefficients $aa$ and $bb$ are defined
as%
\begin{equation}
aa=\frac{1}{2}\left[  3-\frac{1}{v_{0}^{2}}\left(  \frac{1}{\delta}%
+\alpha \right)  \right]  \approx \frac{1}{2}\left(  3-\frac{\alpha}{v_{0}^{2}%
}\right)  >0, \label{e132}%
\end{equation}%
\begin{equation}
bb=\frac{1}{2v_{0}}\left(  \frac{\Omega^{2}c^{4}}{\delta \, \,c_{s}^{4}}%
-\frac{H^{2}}{4}\right)  . \label{e133}%
\end{equation}
The nonlinearity coefficient $aa>0$ is a definite positive because $\alpha
\leq1$, $v_{0}\geq1$. The KdV equation (\ref{e131}) gives a shock solution
instead of soliton if dispersion coefficient $bb$ vanishes, which could happen
at $H=2\Omega c^{2}/\left(  \sqrt{\delta}c_{s}^{2}\right)  $. After some
simplification, the shock condition can be written as follows
\begin{equation}
bb=0\Rightarrow m_{e}m_{i}\lambda_{e}^{2}v_{A}^{2}=\frac{\alpha \hbar^{2}}{12},
\label{e134}%
\end{equation}
It should be noted here that when the dispersion coefficient $bb$ vanishes,
the leading dispersion contribution appears at a higher order, including a
fifth-order spatial derivative term, yielding the Kawahara equation (Kawahara
1972), a possible application of the obtained shock condition.

The limiting cases of KdV equation (\ref{e131}) for magnetosonic waves in a
quantum plasma as discussed here: in dilute plasma case $e^{\beta \mu_{(0)}}%
\ll1$, we have%

\begin{equation}
\frac{\partial v_{ex1}}{\partial \tau}+\frac{(2u_{m}^{2}+v_{A}^{2})}{2u_{m}%
^{2}}\text{ }v_{ex1}\frac{\partial v_{ex1}}{\partial \xi}+\frac{\omega_{pi}%
^{2}}{2u_{m}c_{s}^{3}}\left(  \lambda_{e}^{2}v_{A}^{2}-\frac{\hbar^{2}%
}{12m_{e}m_{i}}\right)  \text{ }\frac{\partial^{3}v_{ex1}}{\partial \xi^{3}}=0,
\label{e135}%
\end{equation}
where magnetosonic and ion-acoustic waves speeds are defined as $u_{m}%
=\sqrt{v_{A}^{2}+c_{s}^{2}}$ and $c_{s}=\sqrt{\kappa_{B}T/m_{i}}$
respectively. Equation (\ref{e135}) is the same as Eq. (30) of Ref. (Ohsawa
and Sakai 1987), when \ (\ref{e135}) is re-written in dimensional form and
replacing $v_{ex1}$ in terms of the first order perturbed electron density
$n_{e1}$ using relation (\ref{e126}), and also neglecting quantum diffraction
effects in the model.

The obtained shock condition ($bb=0$) in the dilute plasma case is
\begin{equation}
\frac{\Omega_{e}}{\omega_{pe}}=\frac{\hbar \omega_{pe}}{2\sqrt{3}\,m_{e}c^{2}%
}\,. \label{e136}%
\end{equation}
or
\begin{equation}
\frac{n_{0}}{B_{0}}=1.49\times10^{29}\, \mathrm{m}^{-3}\,{\mathrm{T}}^{-1}\,,
\label{e137}%
\end{equation}
in terms of SI units. The shock condition in a dilute plasma depends only on
plasma density and magnetic field intensity as long as $T\ll T_{F}$ holds. The
equality is quite applicable to laboratory or astrophysical plasma situations.

However, in a fully degenerate plasma case ($e^{\beta \mu_{(0)}}\ll1$), the KdV
equation (\ref{e131}) gives
\begin{equation}
\frac{\partial v_{ex1}}{\partial \tau}+\frac{(8u_{m}^{2}+v_{A}^{2})}{6u_{m}%
^{2}}\text{ }v_{ex1}\frac{\partial v_{ex1}}{\partial \xi}+\frac{\omega_{pi}%
^{2}}{2u_{m}c_{s}^{3}}\left(  \lambda_{e}^{2}v_{A}^{2}-\frac{\hbar^{2}%
}{36m_{e}m_{i}}\right)  \text{ }\frac{\partial^{3}v_{ex1}}{\partial \xi^{3}}=0,
\label{e138}%
\end{equation}
where $c_{s}=\sqrt{2\varepsilon_{F}/(3m_{i})}$ is the ion-acoustic speed in a
quantum plasma and $H=\hslash \omega_{pe}/(2\kappa_{B}T_{F})$. The obtained
shock condition in a fully degenerate plasma case is given by
\begin{equation}
\frac{\Omega_{e}}{\omega_{pe}}=\frac{\hbar \omega_{pe}}{6\,m_{e}c^{2}}\,,
\label{e139}%
\end{equation}
or
\begin{equation}
\frac{n_{0}}{B_{0}}=2.58\times10^{29}\, \mathrm{m}^{-3}\,{\mathrm{T}}^{-1}\,.
\label{e140}%
\end{equation}
The traveling wave solution of KdV Eq. (\ref{e131}) for one soliton is given
by
\begin{equation}
v_{ex1}=\Psi_{0}\, \mathrm{sech}^{2}\left(  \frac{\eta}{\Delta}\right)  .
\label{e142}%
\end{equation}
where $\eta=\xi-u_{0}\tau$, is the transformed coordinate in the co-moving
frame and $u_{0}$ is the soliton velocity. A single pulse soliton solution is
obtained by applying decaying boundary conditions i.e., $v_{ex1}$, $\partial
v_{ex1}/\partial \eta$ and $\partial^{2}v_{ex1}/\partial \eta^{2}$
$\rightarrow0$ as $\eta \rightarrow \pm \infty$. Here $\Psi_{0}$ is the height
and $\Delta$ is the width of magnetosonic soliton, which are defined as%
\begin{equation}
\Psi_{0}=\frac{3u_{0}}{aa},\text{ \  \  \  \  \  \  \ }\Delta=\sqrt{\frac{4\text{
}bb}{u_{0}}}. \label{e143}%
\end{equation}

It is can be seen from dispersive coefficient $bb$, which contains quantum
diffraction effects defined in Eq.(\ref{e132}), both hump (bright) or dip
(dark) magnetosonic soliton are possible depending on the sign of $bb$. The
height \ of the soliton contains nonlinearity coefficient $aa$ which depends
on quantum degeneracy effects only and remains positive. On condition
$H^{2}/4<(\Omega^{2}c^{4}/\delta c_{s}^{4})$, the dispersive coefficient
$bb>0$ holds for which soliton velocity $u_{0}>0$ remains valid for real
positive values of the soliton width. Therefore, a bright magnetosonic soliton
moving with supersonic speed is formed. However, if the dispersive coefficient
has negative values i.e., $bb<0$, which could happens at $H^{2}/4>(\Omega
^{2}c^{4}/\delta c_{s}^{4})$ then soliton velocity $u_{0}<0$ should hold for
the validity of the soliton solution (\ref{e142}) (Belashov and Vladimirov
2005; Haas and Mahmood, 2015;2016;2018). Therefore at these conditions there
is possibility of dark soliton moving with subsonic speed. The variation of
dispersive coefficient $bb$ with thermal temperature is plotted at different
magnetic field intensities in Fig.5 for arbitrary degenerate electrons
plasmas. It is evident from the numerical plot that the dispersive coefficient
$bb<0$ holds for electron temperatures of the order of magnitude $T>10^{6}$
\textrm{K} and dark magnetosonic solitons moving with subsonic speed are
formed. The range of electron temperature for the formation of dark soliton
structure is reduced with increase in the magnetic field intensity as shown in
the figure.


\begin{figure}[tbh]
\begin{center}
\includegraphics[width=8cm]{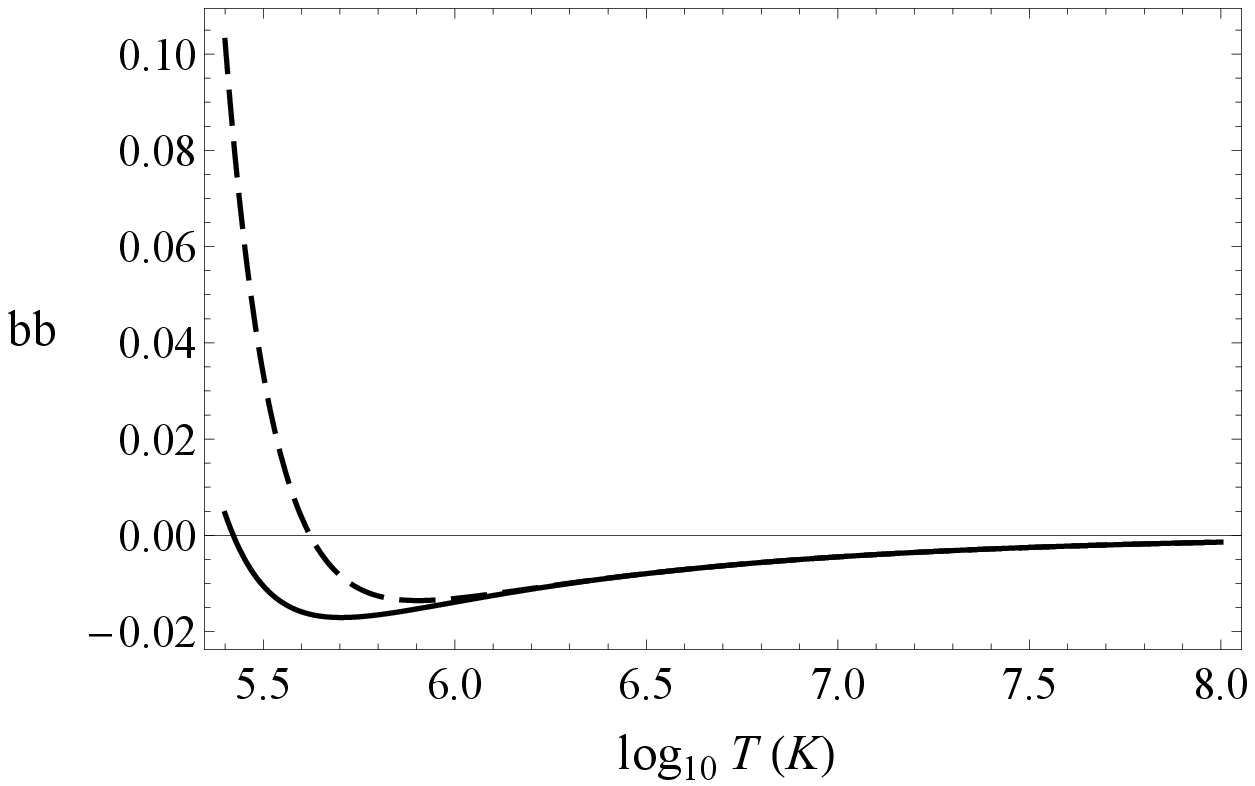}
\end{center}
\caption{Dispersive coefficient $bb$ \ of KdV equation for magnetosonic wave
is shown as a function of electron temperature from Eq. (\ref{e133}) for
$B_{0}=3$ {$\mathrm{T}$} (lower, solid curve) and $B_{0}=6$ {$\mathrm{T}$}
(upper, dotted curve) at $z=1$. }%
\label{figure5}%
\end{figure}

\section{Numerical Estimates and Applications}

In this section, the numerical estimates of the physical dense parameters will
be found out for the propagation of linear and nonlinear waves in arbitrary
degenerate electron plasma. The material reviews the presentation in (Haas and
Mahmood 2018). According to our general presented theory of arbitrary
degeneracy of electrons, which lies in the intermediate range where the
thermal and Fermi temperature are almost equal i.e.,%
\begin{equation}
T=T_{F}\,, \label{e144}%
\end{equation}
where electrons Fermi temperature $T_{F}=\varepsilon_{F}/\kappa_{B}$ is
defined in terms of Fermi energy.

In order to find numerical estimates of the fugacity $z=$ $e^{\beta \mu_{(0)}}%
$and coefficient $\alpha$ for the intermediate range, the relation
(\ref{e801}) can be re-written (Eliasson and Shukla 2008) as follows,%

\begin{equation}
\mathrm{Li}_{3/2}(-z)=-\frac{4}{3\sqrt{\pi}}\,(\beta \varepsilon_{F})^{3/2}\,,
\label{e145}%
\end{equation}
Accordingly intermediate dense plasmas in which electron thermal and Fermi
temperature are equal, therefore $\beta \varepsilon_{F}=1$ for which
equilibrium fugacity comes out to be $z=0.98$ from Eq.(\ref{e145}) and
$\alpha=0.80$ from Eq.(\ref{e293}). It is also evident from Fig.2 that
$\alpha=0.80$ lies in the intermediate dilute-degenerate plasma situation.

The two phenomenon such as quantum degeneracy through electron Fermi pressure
and quantum diffraction effects due to wave like nature of electrons are
included in the model. The quantum diffraction effects gives extra dispersion
of linear wave in dense plasmas and modified the width of the soliton
structure. The quantum diffraction parameter $H$ plays an important role in
the formation of bright or dark soliton in magnetized dense plasmas.
Theoretically speaking the dark soliton are formed for the large values of
quantum diffraction parameter $H$ which in reality cannot be increased with
limits under the present ideal Fermi-gas model. In fact for large values of
$H$ one would enter to strongly plasma regime which is not applicable to the
present model. Therefore in order to find the validity range of quantum
diffraction parameter $H$, the coupling parameter $g$ is analyzed such that
$g=l/r_{s}$, and
\begin{equation}
l=-\frac{e^{2}}{12\pi \varepsilon_{0}\kappa_{B}T}\, \frac{n_{0}\Lambda_{T}^{3}%
}{\mathrm{Li}_{5/2}(-z)}\text{.} \label{e146}%
\end{equation}
Here $l$ is a generalized Landau length (Kremp et al. 2005; Ichimaru 2004;
Lifshitz 1981) involving the thermal de Broglie wavelength $\Lambda_{T}%
=[2\pi \hbar^{2}/(m_{e}\kappa_{B}T)]^{1/2}$, and $r_{s}=(4\pi n_{0}/3)^{-1/3}$
is the Wigner-Seitz radius.

In the dilute plasma case, one has $e^{2}/(4\pi \varepsilon_{0}\,l)=(3/2)\kappa
_{B}T$, so that $l$ would be the classical distance of closest approach in a
binary collision, for average kinetic energy. The degeneracy effects have been
incorporated in the mean kinetic energy as described in (\ref{e146}). It could
be noticed that indiscriminate increase in the quantum value of $H$ enhances
the non-ideality effects such as bound states and dynamical screening (Kremp
et al. 2005). Therefore it can be concluded that the dark soliton and shock
wave solutions of KdV and ZK equations, which theoretically needs large values
of quantum diffraction parameter i.e., $H>2$ lies in the outside range of
ideal (or less collisional) dense plasma situation i.e., $g<1$. However, there
is still possibility that wave nature of electrons associated with quantum
diffraction parameter provides important corrections for the least reasonable
values of $H$.

Now in order to find the estimates of the range of minimum to maximum
wavelengths of the wave as a function of electron Fermi energy in the
intermediate-dilute degenerate plasma, for the consistency of our present
investigation, we have from Eq. (27) of (Haas and Mahmood 2015) i.e.,
\begin{equation}
k_{\mathrm{min}}\equiv \frac{2\sqrt{3}m_{e}c_{s}}{\hbar}\ll k\ll \frac
{\omega_{pi}}{c_{s}}\equiv k_{\mathrm{max}}\,. \label{e147}%
\end{equation}
The above equation has been obtained under long wavelength assumption from
fluid model and static response of electrons. In case of hydrogen plasma and
$T=T_{F}$, we have $k_{\mathrm{max}}>k_{\mathrm{min}}$ for $n_{0}%
<9.81\times10^{35}\, \mathrm{m}^{-3}$ (solid density) from Eq. (\ref{e147}).
The dense plasma density lies in the nonrelativistic regime, according to our
present study. The first in equality in Eq. (\ref{e147}) can be ignored if the
quantum degeneracy effects are more dominant then the quantum diffraction
effects. The suitable range of wavelength $\lambda_{\mathrm{min}}%
=2\pi/k_{\mathrm{max}}\ll \lambda \ll \lambda_{\mathrm{max}}=2\pi/k_{\mathrm{min}%
}$ as a function of electron energy for intermediate dilute degenerate
hydrogen plasma as a function of electron Fermi energy is plotted in
Fig.6\textbf{, }which lies in the nanometric scale from extreme ultraviolet to
soft X-rays. However, in case of high densities exist in compact astrophysical
objects like white dwarfs and neutron stars, electron Fermi momentum becomes
comparable to $m_{e}c$ and dense plasma needs relativistic treatment which
beyond the scope of our present studies.


\begin{figure}[tbh]
\begin{center}
\includegraphics[width=8cm]{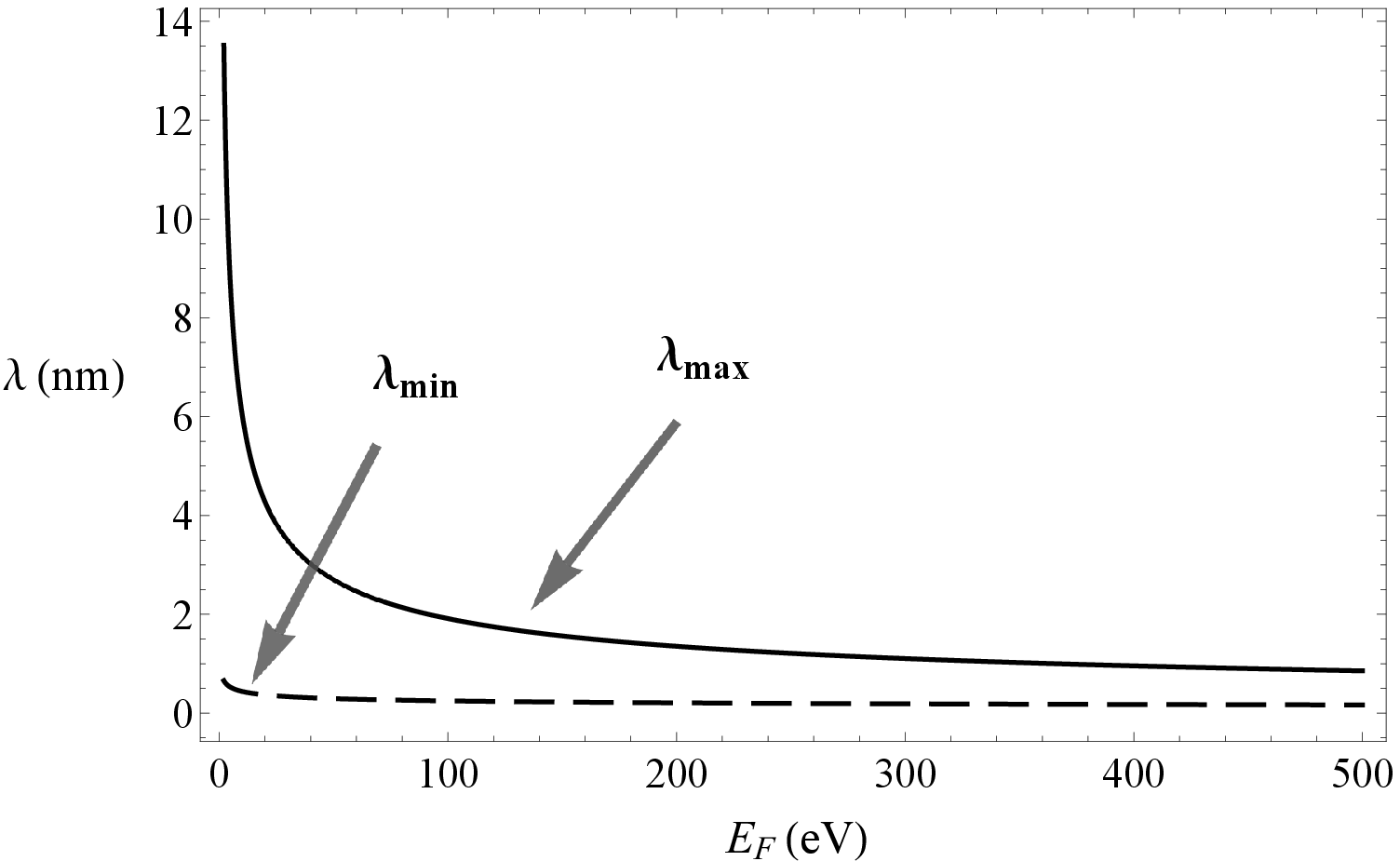}
\end{center}
\caption{Upper, continuous curve: maximum wavelength $\lambda_{\mathrm{max}}$;
lower, dashed curve: minimal wavelength $\lambda_{\mathrm{min}}$, consistent
with Eq. (\ref{e147}) are plotted as a function of the electronic Fermi energy
$\varepsilon_{F}$ in $\mathrm{eV}$, for hydrogen plasma in the intermediate
dilute-degenerate regime where $T=T_{F}$ holds. }%
\label{figure6}%
\end{figure}

In order to justify the cold and nondegenerate ideal ionic fluid and avoiding
the strong ion coupling effects, the necessary condition for ionic coupling
parameter (Kremp et al. 2005) is defined as%

\begin{equation}
g_{i}=\frac{e^{2}}{4\pi \varepsilon_{0}r_{s}}\, \left(  \frac{3\kappa_{B}T_{i}%
}{2}\right)  ^{-1}\ll1\, \label{e148}%
\end{equation}
where $T_{i}$ is the the ion thermal temperature and $T_{i}<T_{F}$ should
hold. It is believed that for $g_{i}\approx172$ in one-component plasma the
ionic liquid or crystallization start to happen (Murillo 2004). The allowable
region between the two straight lines $T_{i}=T_{F}$ and $g_{i}=1$ for cold and
weakly coupled ions is shown in Fig.7. The minimum number density of cold and
weakly coupled ions comes out to be $n_{0}=3.63\times10^{28}\, \mathrm{m}%
^{-3}$ which lies in the range of typical solid density plasmas.


\begin{figure}[tbh]
\begin{center}
\includegraphics[width=8cm]{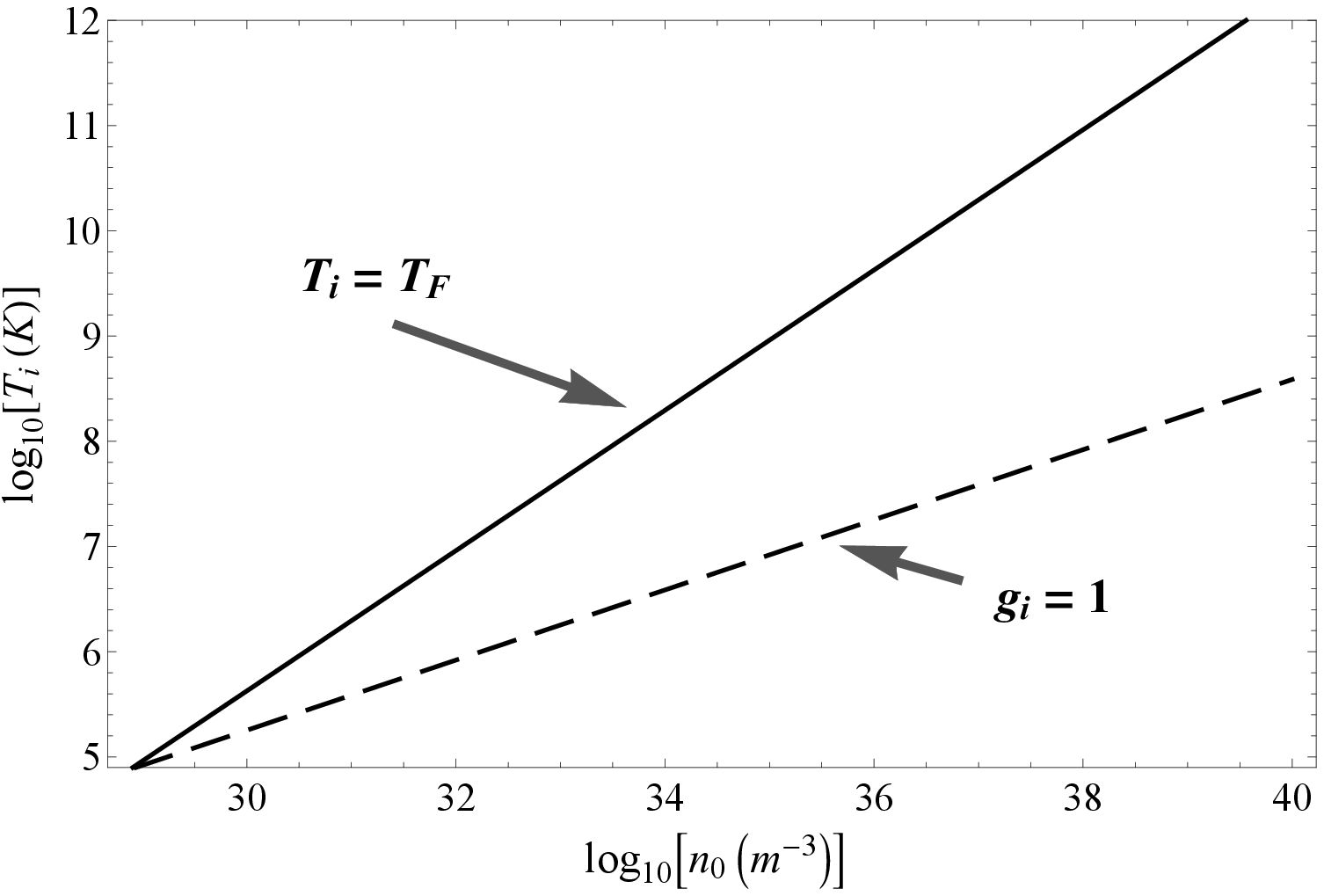}
\end{center}
\caption{The cold and weakly coupled ions region below the upper, continuous
straight line (where the ionic temperature $T_{i}$ equals the electronic Fermi
temperature $T_{F}$) and above the lower, dashed straight line (where the
ionic coupling parameter $g_{i}=1$) is shown using Eq. (\ref{e148}). }%
\label{figure7}%
\end{figure}

\section{Conclusion}

The main findings our present investigation to study the linear and nonlinear
ion-acoustic waves and magnetosonic waves in a quantum plasma with arbitrary
degeneracy of electrons. The equation of state for arbitrary degeneracy of
electrons is found out in the form of polylogarithms obtained by using
Fermi-Dirac integral, which is a function of both chemical potential and
temperature. It's limiting cases for dilute and dense plasma regions are also
discussed. A dimensionless parameter ($\alpha$), which involves dimensionality
and temperature, is defined in front of quantum force in electron momentum
quantum fluid equation to fit the results obtained from quantum kinetic theory
in the long wavelength limit. A mathematical relation of $\alpha$ in the form
of polylogarithms function of equilibrium chemical potential and temperature
is obtained its numerical values are found out in the limiting cases for
dilute and dense plasmas. The KdV equation for ion-acoustic waves in
unmagnetized quantum plasma with arbitrary degeneracy of electrons with its
soliton solution is obtained. The bright IAW soliton in unmagnetized quantum
plasma is formed for the quantum diffraction values $H<2$ (moving with
supersonic speed), while dark soliton structure is formed for $H>2$ (moving
with subsonic speed). The ion-acoustic wave dispersion effects in unmagnetized
quantum plasma appears only through quantum diffraction effects from Bohm potential.

The ZK equation for two dimensional propagation of ion-acoustic soliton is
also obtained in a magnetized quantum plasma with arbitrary degeneracy of
electrons. It is found that wave dispersion effects appear through quantum
diffraction parameter ($H$) and ion Larmour radius effect in the perpendicular
direction of the magnetic field. The conditions for existence of bright or
dark ZK ion-acoustic solitons in a magnetized quantum plasma with quantum
diffraction parameter and magnetic field intensity are discussed in detail.
Similarly, the KdV equation for weakly nonlinear magnetosonic waves in a
quantum plasma with finite temperature effects of electrons is also studied.
It is found that both bright or (dark) magnetosonic soliton structures are
possible moving with supersonic and (subsonic) speeds in a magnetized quantum
plasma depending on the plasma density, electron temperature and magnetic
field intensity. Other than solitons, the shock wave solution with its
conditions for formation depending on the values of quantum diffraction
parameter and magnetic field intensity are also discussed. Also, a general
coupling parameter in the form of polylogarithms is proposed and conditions
for ideal (or collisionless) dilute or dense plasmas are worked out. The
mathematical expressions of minimal criteria for the electron temperature and
maximal quantum diffraction parameter on the formation of dark solitons for
the ideality and weak coupling system having arbitrary degeneracy of electrons
are also discussed and their physical conditions for real systems are also
pointed out. The physical parameters of hydrogen plasmas in the intermediate
dilute-degenerate limits where $T=T_{F}$\ holds for fully ionized electron
case with arbitrary temperature of degenerate electrons are also presented. It
is found that for hydrogen plasmas (solid-density) with arbitrary degeneracy
of electrons, the wavelength ($\sim$ nanometer scale) of plasma waves lies in
the range from extreme ultraviolet to soft x rays. Therefore, we are hopeful
that the physical parameters developed form our present theory will be useful
for experimental and observational verifications of linear and nonlinear
ion-acoustic and magnetosonic waves in laboratory and in nature with large
range of electron degeneracy regime.

\textbf{Acknowledgments:} F.~H.~ acknowledges the support by Con\-se\-lho
Na\-cio\-nal de De\-sen\-vol\-vi\-men\-to Cien\-t\'{\i}\-fi\-co e
Tec\-no\-l\'o\-gi\-co (CNPq).

\vspace{.5cm}

{\it This version is faithful (unconstrained by unfair refereeing).}

\bigskip

\textbf{Appendix:} \textbf{Functions defined after collecting first and second
order terms from perturbation theory}

The functions $f_{1}$ to $f_{9}$ defined in the equation (\ref{e123}) are
given as follows:%

\begin{equation}
f_{1}=\frac{\partial n_{i1}}{\partial \tau}+\frac{\partial}{\partial \xi}\left(
n_{i1}v_{ix1}\right)  \,,\text{ \  \  \  \  \  \  \  \  \  \  \  \  \  \  \  \  \  \ }%
f_{2}=\frac{\partial n_{e1}}{\partial \tau}+\frac{\partial}{\partial \xi}\left(
n_{e1}v_{ex1}\right)  \,, \tag{a1}%
\end{equation}%
\begin{equation}
f_{3}=\frac{\partial v_{ix1}}{\partial \tau}+v_{ix1}\frac{\partial}{\partial
\xi}v_{ix1}-\Omega \text{ }v_{iy1}B_{z1}\,,\text{ \  \  \  \  \  \  \  \  \  \  \  \ }%
f_{4}=-v_{0}\frac{\partial v_{ix1}}{\partial \xi}+\Omega \text{ }v_{ix1}%
B_{z1}\,, \tag{a3}%
\end{equation}%
\begin{equation}
f_{5}=\frac{\partial v_{ex1}}{\partial \tau}+v_{ex1}\frac{\partial}{\partial
\xi}v_{ex1}+\delta \, \Omega \text{ }v_{ey1}B_{z1}-\delta \, \alpha \text{ }%
n_{e1}\frac{\partial n_{e1}}{\partial \xi}-\delta \, \frac{H^{2}}{4}\text{
}\frac{\partial^{3}n_{e1}}{\partial \xi^{3}}\,, \tag{a5}%
\end{equation}%
\begin{equation}
f_{6}=-v_{0}\frac{\partial v_{ey1}}{\partial \xi}+\Omega \text{ }v_{ex1}%
B_{z1}\,,\text{ \  \  \  \  \  \  \ }f_{7}=-\Omega \frac{\partial B_{z1}}%
{\partial \tau}\,, \tag{a6}%
\end{equation}%
\begin{equation}
f_{8}=\left(  n_{e1}-n_{i1}\right)  v_{ex1}\,,\text{
\  \  \  \  \  \  \  \  \  \  \  \  \  \  \  \  \  \  \  \  \  \  \  \  \  \ }f_{9}=\frac{c_{s}^{2}%
}{c^{2}}\left(  n_{e1}v_{ey1}-n_{i1}v_{iy1}\right)  \,. \tag{a8}%
\end{equation}

\end{document}